\documentclass[sigconf,nonacm]{acmart}
\usepackage[T1]{fontenc}
\usepackage{graphicx}

\usepackage[caption=false,font=footnotesize]{subfig}
\usepackage{xspace}
\usepackage{xcolor}
\usepackage{multirow}
\usepackage{amsmath}
\usepackage{enumitem}
\usepackage{bm}

\usepackage{booktabs}
\usepackage{tabularx}
\usepackage{pifont}
\usepackage{hyperref}
\usepackage{cleveref}
\crefname{figure}{Fig.}{Figs.}
\Crefname{figure}{Fig.}{Figs.}
\usepackage{xurl}
\usepackage{stfloats}
\usepackage{cuted}
\usepackage{capt-of}

\usepackage{color}

\usepackage{tikz}
\usetikzlibrary{calc}

\newcommand{\CTTF}{\textsc{CttF}\xspace}

\newcommand{\good}{\textcolor{green}{\ding{51}}}
\newcommand{\bad}{\textcolor{red}{\ding{55}}}
\newcommand{\somewhat}{\textcolor{orange}{\ding{51}}}

\newcommand{\dottedcircle}{%
  \tikz[baseline=-0.6ex]{
    \draw[densely dotted,line width=1pt] (0,0) circle (0.6ex);
  }%
}

\newcommand{\missing}{\textcolor{red}{\dottedcircle}}

\newcommand{\unknown}{\textcolor{gray}{\textbf{?}}}

\newcommand{\mypara}[1]{\vspace{1.5pt}\noindent{\bf {#1.}}}
\newcommand{\myparait}[1]{\vspace{1.5pt}\noindent{\emph {#1.}}}

\newcommand{\esoric}[1]{#1}
\newcommand{\wpes}[1]{#1}

\begin{document}
\title{Cache to the Future: A Distributed Webpage Archive for Internet Blackouts}
\author{Ross Evans}
\orcid{0009-0008-6484-011X}
\affiliation{%
  \institution{University of Waterloo}
  \country{Waterloo, ON, Canada}
}
\email{rpevans@uwaterloo.ca}

\author{Diogo Barradas}
\orcid{0000-0003-0338-2692}
\affiliation{%
  \institution{University of Waterloo}
  \country{Waterloo, ON, Canada}
}
\email{diogo.barradas@uwaterloo.ca}

\settopmatter{printacmref=false} %
\renewcommand\footnotetextcopyrightpermission[1]{} %
\acmDOI{} %
\acmISBN{} %

\begin{abstract}
Internet blackouts, occurring due to technological mishaps or intentional governmental action, prevent citizens from accessing the internet.
Citizens in regions where internet blackouts are common have utilized blackout-resistant technologies to maintain communication.
Such technologies often rely on mobile mesh networks to provide limited messaging services.
However, no technology currently exists which can provide continued access to knowledge sources on the web during a blackout.

We present Cache to the Future (\CTTF): a system to cache and deliver static content hosted on the web during a blackout.
\CTTF's distributed community ratings crowdsources caching at scale while cryptographic constructs (digital signatures, proofs-of-work) mitigate adversarial interference.
Our realistic simulations demonstrate \CTTF delivering content at city-scale across a wide range of benign and adversarial scenarios.

\end{abstract}



\maketitle              %

\section{Introduction}

Internet blackouts prevent citizens within an affected region from accessing internet services~\cite{bischof2023destination}.
While not all internet blackouts are malicious, as may occur due to natural disasters or other technological failures~\cite{aceto2018comprehensive,rogersoutage}, repressive governments have often instated blackouts intentionally via technical means during political unrest, protests, and elections~\cite{accessnow,tagat2024net}.
Today, state-instated blackouts are a growing issue: in 2024, Access Now documented 296 blackouts in 54 countries: the highest number of blackouts since measurement began in 2016~\cite{accessnow}.

To counter the lack of internet access during blackouts, citizens often rely on \textit{blackout-resistant technologies}. These technologies aim to re-establish electronic communications via alternative mediums and protocols outside the internet~\cite{hongkongbridgefy,sharma2023dolphin,hammas2025evading}, and
have found applications in disaster scenarios~\cite{lu2017teamphone,bluemergency,10.1145/3379503.3403532,10.1145/2989250.2989257} as well as during political protests in multiple countries~\cite{ukraineblackout,myanmarblackout,afghanistanblackout,firechat}.
Most of these technologies rely on mobile mesh-networks to facilitate communication~\cite{perry2022strong,pradeep2022moby,inyangson2024amigo,kamali2024anix}, using short-range wireless technologies such as Bluetooth and Wi-Fi Direct to exchange information in a peer-to-peer fashion.

Unfortunately, mobile-mesh networks pose a challenging routing environment which prior works have struggled to adapt to.
The short-range nature of these technologies coupled with natural human movement results in a rapidly evolving network topology with low density and brief connectivity. 
Traditional routing algorithms are ineffective in this setting; hence, most prior works in this space make use of epidemic routing~\cite{inyangson2024amigo} (also known as ``flooding'' or ``gossiping'')~\cite{vahdat2000epidemic}.
Yet, besides imposing significant bandwidth requirements, Inyangson et al.~\cite{inyangson2024amigo} revealed that blackout-resistant tools using epidemic routing suffer from poor delivery rates due to congestion at the physical layer (L1), defying earlier beliefs about the scalability of such tools.

Epidemic routing is also prone to interference in adversarial scenarios; a state-level adversary wishing to disable the blackout-resistant technology 
can mount a denial-of-service attack through coordinated spamming~\cite{lerner2016rangzen}.
With ephemeral user connections and Bluetooth's low data-transfer rates, spammed adversarial messages can prevent users from relaying benign messages in their queues.
Various strategies to combat such attacks have been explored~\cite{pradeep2022moby,kamali2024anix,lerner2016rangzen}, but nearly all of them require trust establishment between users, which can be difficult to achieve at scale in practical usage scenarios. Within repressive environments, pervasive fear of surveillance and betrayal can erode interpersonal trust to the point that users are unwilling to rely on each other at all~\cite{xue2024bridging}, posing fundamental challenges to trust-based blackout-resistant technologies.

In addition to routing challenges, prior blackout-resistant technologies have largely concentrated on providing messaging and microblogging services in the event of a blackout~\cite{inyangson2024amigo,kamali2024anix,bienstock2023asmesh}.
\esoric{However, blackouts also disrupt access to web-based knowledge, including reference sources, medical information, and community webpages.}
In previous studies on the perceived impacts of internet outages~\cite{grandhi2020internet,lupien2017wait}, respondents noted that resources like Wikipedia made them feel more informed, and feared feeling less capable if unable to access them.
Beyond unintentional blackouts, information access is also critical in censored areas; a survey %
identified access to ``work-related resources'' and ``educational materials like Wikipedia'' as key motivators for circumventing censorship~\cite{xue2024bridging}. 
Some circumvention tools disseminate static content: e.g., the \textit{Uncensored Library}~\cite{uncensoredlibrary} hosts \esoric{a fixed set of} censored articles and \textit{Collateral Freedom}~\cite{rsf2024collateral} mirrors censored media websites.
Yet, despite many innovative approaches~\cite{khattak2016sok,tschantz2016sok}, circumvention tools still rely on an active internet connection. During a blackout, however, internet access is revoked entirely, and existing blackout-resistant technologies fail to deliver rich and information-dense web content at scale.

To address this gap, we focus on enabling users to retain access to webpages even when no internet is available. In particular, we aim to deliver websites with largely static content which can be fetched without requiring user logins or persistent state. In contrast to webpages such as social media feeds, which serve dynamic content and depend on real-time user interactions with backend services, the static content hosted by knowledge sources is amenable to pre-fetching and caching for later access.

We present Cache to the Future (\CTTF), a system that enables access to information during a blackout via cached webpages.
\CTTF operates in a \textit{pre-blackout phase} and a \textit{blackout phase}.
In the pre-blackout phase, users can cache websites and rate them according to their perceived usefulness. During this period, \CTTF users can interact with each other via short-range connections and accumulate local averages of the ratings received from other users.
\CTTF users cache pages according to these local averages: pages with higher ratings are cached across more devices to improve accessibility.
Communities with high technological or political engagement can thereby seed useful webpages across the devices of less engaged users to establish a distributed internet archive.
The crowdsourced nature of \CTTF distributes the high storage requirements of this archive across many devices and allows users to fetch content they may have not previously seen.
During the blackout phase, users request pages and have their requests fulfilled by other users who have cached them.
Requests are stored until a short-range connection with a user who has cached the requested page can be established.
\CTTF avoids the unfeasible bandwidth overheads of epidemic routing in this context while effectively serving pages with rich content (CSS, JS, images).

Beyond internet outages caused by natural disasters or infrastructure failures, \CTTF also considers hostile scenarios where a state-level censor aims to disrupt the system's activity. %
\CTTF limits an adversary's ability to deploy Sybil nodes and manipulate caching by leveraging a proof-of-work scheme during page rating exchanges. It also allows users to prove page contents and provenance by fetching pages during the pre-blackout phase via a suite of trusted censorship-circumvention proxies which sign them. In case all proxies are inaccessible, \CTTF can fetch multiple copies of the page from independent users and perform a ``majority vote''. Finally \CTTF offers resilience against localized signal jamming through opportunistic data exchanges; even if certain areas are jammed, users can eventually encounter others---outside the jammed zones---who carry the desired content. We show this robustness through a comprehensive set of simulations in a city-wide scale (\S\ref{sec:evaluation-results}).

To evaluate \CTTF, we implemented an Android prototype and conducted micro-benchmarks, as well as developed simulations based on real-world data to model its performance at a city-wide scale.
This dataset contains 75 days of GPS movement data for 25\,000 individuals in a Japanese city, locating them at 30-minute intervals within one of 40\,000 grid cells with 0.25${\text{km}}^2$ area.
Our evaluation demonstrates that in a benign scenario, users can fetch from a set of 100\,000 pages with 75\% probability across a 2-month blackout.
In addition, the median latency for the 10\,000 most popular pages is under 24 hours.
Furthermore, \CTTF demonstrates strong adversarial resistance: increasing Sybil nodes to 25\% of the network only drops request satisfaction by 5\% in a week-long blackout compared to the benign scenario, and adversaries must jam 2\,500 $\text{km}^2$ to significantly impact \CTTF performance.

\mypara{Contributions} We summarize our contributions as follows:
\begin{itemize}[leftmargin=*]
\item We explore a new design point in blackout-resistant systems by caching web-hosted knowledge sources rather than providing interactive communication platforms.
\item We propose a communication model that avoids the bandwidth overheads of epidemic routing, retrieving data via direct user-to-user exchanges.
\item We develop a \CTTF prototype that supports access to cached webpages during blackouts, while withstanding content manipulation and signal jamming attacks.
\item We evaluate \CTTF through micro-benchmarks and city-scale simulations using realistic, longitudinal mobility traces from a major metropolitan area.
\end{itemize}

\section{Mobile Ad-Hoc Network Communication}

\begin{table*}[t!]
\caption{Prior works evaluated w.r.t. to routing approach, threat model, and testbed.}
\vspace{-3mm}
\label{table:comparison}
\resizebox{0.7\textwidth}{!}{%
\begin{tabular}{llcccccccccc}
\toprule

\textbf{Technology Category} & \textbf{Application} & \multicolumn{5}{c}{\textbf{Routing Approach}} & \multicolumn{3}{c}{\textbf{Attacks}} & \multicolumn{2}{c}{\textbf{Evaluation}} \\
\cmidrule(lr){3-7}
\cmidrule(lr){8-10}
\cmidrule(lr){11-12}
& & ER & BF & CR & T & PMR & DoS & J & ASM & RWD & CS \\
\midrule

\multirow{10}{*}{Messaging} 
  & 1am~\cite{liu2015performance} & \good & \good & \bad & \bad & \bad & \good & \good & \bad & \somewhat & \good \\
  & Amigo~\cite{inyangson2024amigo} & \good & \good & \good & \somewhat & \bad & \bad & \bad & \bad & \bad & \bad \\
  & Anix~\cite{kamali2024anix} & \good & \good & \bad & \good & \bad & \good & \bad & \good  & \bad & \good \\
  & ASMesh~\cite{bienstock2023asmesh} & \good & \bad & \bad & \bad & \bad & \bad & \bad & \bad & \bad & \good\\
  & Briar~\cite{briar} & \good & \unknown & \bad & \bad & \bad & \missing & \missing & \missing & \missing & \missing \\
  & Bridgefy~\cite{bridgefy} & \good & \unknown & \bad & \bad & \bad & \missing & \missing & \missing & \missing & \missing\\
  & NewNode~\cite{newnode} & \good & \unknown & \bad & \bad & \bad & \missing & \missing & \missing & \missing & \missing \\
  & Mirage~\cite{mirage} & \good & \bad & \bad & \bad & \good & \bad & \bad & \bad & \good& \good\\
  & Moby~\cite{pradeep2022moby} & \good & \bad & \bad & \good & \bad & \good & \good & \good & \somewhat & \good \\
  & Perry et al.*~\cite{perry2022strong} & \good & \good & \good & \bad & \bad & \bad & \bad & \bad  & \bad & \bad \\
  & Rangzen~\cite{lerner2016rangzen} & \good & \bad & \bad & \good & \bad & \good & \good & \good & \somewhat & \good \\

\midrule

\multirow{4}{*}{MANET File-Sharing} 
  & BTM~\cite{rajagopalan_cross-layer_2006} & \multicolumn{5}{c}{Proactive route maintenance} & \bad & \bad & \bad & \bad & \bad\\
  & Liu et al.*~\cite{liu_cooperative_2011} & \good & \bad & \bad & \bad & \bad & \bad & \bad & \bad & \somewhat & \bad \\
  & M2MShare~\cite{palazzi_social-aware_2012} & \multicolumn{5}{c}{ORION~\cite{klemm_special-purpose_2003} + Mobility} & \bad & \bad & \bad & \bad & \good \\
  & ORION~\cite{klemm_special-purpose_2003} & \multicolumn{5}{c}{Reactive route generation} & \bad & \bad & \bad & \bad & \bad \\

\midrule

Webpage Access & \CTTF & \multicolumn{5}{c}{Non-Routing} & \good & \good & \good & \good & \good \\
\bottomrule
\end{tabular}}\hspace{1em}\begin{minipage}{0.25\textwidth} %
\footnotesize
\textbf{Legend:} \\
\sloppy
\scriptsize
\textbf{ER:} Epidemic Routing\\
\textbf{BF:} Bloom Filters\\
\textbf{CR:} Clique Routing\\
\textbf{T:} Trust Metrics \\
\textbf{PMR:} Private Mobility-Routing\\
\textbf{DoS:} Denial~of~Service\\
\textbf{J:} Jamming \\
\textbf{ASM:}~App~Specific~Manipulation\\
\textbf{RWD:}~Real~World~Data\\
\textbf{CS:}~City-Scale\\

\somewhat: Only somewhat supported\\
\missing: Not formally defined\\
\unknown: Unknown, lack of documentation\\
*: Author names used when apps are not named
\fussy
\end{minipage}
\end{table*}

This section discusses the landscape of existing blackout-resistant technologies based on ad-hoc networks. As previously mentioned, these works largely focus on messaging and microblogging, essentially targeting a different end goal than that of \CTTF's. Nevertheless, an analysis over these technologies' functionalities remains useful for contrasting with \CTTF's envisioned operational setting.

Beyond the realm of blackout-resistant technologies, previous works have also studied file-sharing in mobile ad-hoc networks (MANETs). These networks lack dedicated infrastructure: nodes may travel and change links to other devices on a frequent basis. Smartphone-based mesh networks are an extreme version of MANETs, with very sparse nodes and brief connection durations~\cite{bang2013manet}. While MANET research is relevant to our context---since file-sharing in such networks shares goals with \CTTF's cached webpage retrieval---the connectivity constraints in smartphone mesh networks are typically too severe for conventional MANET protocols to operate effectively. 
Further, these protocols are not typically designed with adversarial behaviour in mind \cite{hui2008bubble}.

In our analysis (see \Cref{table:comparison}), we compare prior works across their routing protocols, threat models, and evaluation testbeds to inform our design choices for \CTTF.

\subsection{Routing Approaches}
We first detail the routing approaches employed by blackout-resistant technologies followed by the approaches of MANET file-sharing technologies.
The MANET file-sharing works generally assume a network path exists between origin and destination, while the blackout-resistant communication technologies assume a largely disconnected network.

\mypara{Epidemic Routing}
All of the blackout-resistant messaging applications we survey rely on some form of epidemic routing.
When two nodes come in contact, they share their own messages and forward messages previously received from other users.
Epidemic routing can effectively propagate messages but requires significant bandwidth to function.
Nodes often waste bandwidth re-transmitting messages that have already been received via alternate paths.

\mypara{Bloom Filters} 1am~\cite{liu2015performance}, Amigo~\cite{inyangson2024amigo}, Anix~\cite{kamali2024anix}, and Perry et al.~\cite{perry2022strong} all consider using Bloom filters to reduce bandwidth.
Bloom filters allow users to send others a short digest of their stored messages.
The digest can help prevent users forwarding messages the other user has already received.
Briar~\cite{briar}, Bridgefy~\cite{bridgefy}, and NewNode~\cite{newnode} are each non-academic works without technical papers published detailing their implementations; we lack sufficient data to state whether they employ Bloom filters.

\mypara{Clique Routing} In dense, geographically-limited protest scenarios, Perry et. al.~\cite{perry2022strong} introduced ``static clique routing'', where designated clique leaders exchange messages on behalf of their clique to reduce bandwidth.
Amigo~\cite{inyangson2024amigo} extended this with a ``dynamic clique routing'' approach where new clique leaders are elected per epoch to account for shifting crowd mobility.
Unfortunately, Amigo also demonstrated that epidemic routing in all of its forms---including Bloom filter and both clique variants---suffers from real-world performance issues in densely-packed scenarios due to excessive collisions in the network's physical layer~\cite{inyangson2024amigo}.

\mypara{Trust}
Anix~\cite{kamali2024anix}, Moby~\cite{pradeep2022moby}, and Rangzen~\cite{lerner2016rangzen} use trust metrics in message propagation.
Rangzen and Moby both privately compute shared contacts between peers and prioritize which messages are exchanged accordingly.
Anix also relies on social contacts while additionally providing the ability to see how many trusted contacts upvoted or downvoted a message.

\mypara{Private Mobility Routing} Mirage~\cite{mirage} forwards messages via differentially-private mobility graphs.
Users broadcast selectively to reduce overhead while increasing the chance that messages are routed to their intended destination.
Mirage assume users know the approximate physical location of their messages' addressees.

\mypara{MANET File-Sharing} In contrast to the blackout-resistant technologies, most MANET file-sharing applications assume a path between source and destination when serving a request.
ORION~\cite{klemm_special-purpose_2003} and BTM~\cite{rajagopalan_cross-layer_2006} optimize ad-hoc internet routing with node mobility via overlay networks which generate and maintain routes.
M2MShare~\cite{palazzi_social-aware_2012} builds upon ORION; they improve routing performance by collecting movement patterns of users.
Tracking movement, however, poses a privacy risk in a blackout scenario~\cite{hasan2013building}.
Liu et al.~\cite{liu_cooperative_2011} studied a setting where a portion of nodes have internet access and wish to provide file downloading services for the non internet-connected nodes in the network.
But, they also utilize epidemic routing in a densely-populated scenario which, as previously mentioned, suffers from performance issues. Nodes are also required to collect frequent contacts for file metadata discovery which poses privacy concerns.

\vspace{1mm} \hspace{-5mm}
\fbox{\parbox {0.97 \linewidth} {\textbf{Design goal 1.} \CTTF must use a communication model which functions effectively at a city-wide scale and resists adversaries.}}

\subsection{Attacks Considered in the Threat Model}
\label{subsec:considered-attacks}

We now present the attacks considered by previous works. 
The MANET file-sharing systems are not designed for adversarial blackouts and do not consider any attacks.
Briar~\cite{briar}, Bridgefy~\cite{bridgefy}, and NewNode~\cite{newnode} do not define formal threat models nor demonstrate attack resistance.
Mirage~\cite{mirage} only considers adversarial inference of mobility patterns from routing.
For the remaining works, we highlight common attacks on system performance.

\mypara{Denial-of-Service} All the blackout-resistant apps we survey assume the adversary may operate Sybil nodes in the mesh-network which can behave maliciously to degrade system performance.
1am~\cite{liu2015performance}, Anix~\cite{kamali2024anix}, Moby~\cite{pradeep2022moby}, and Rangzen~\cite{lerner2016rangzen} simulate attacks where the adversary refuses to forward or respond to legitimate content and instead floods the network with their own spam data.

\mypara{Jamming}
The blackout-resistant messaging apps we survey use short-range communication to form the mesh network and are susceptible to jamming attacks.
These attacks assume the adversary can prevent Bluetooth connections from being formed in a defined area.
1am~\cite{liu2015performance}, Moby~\cite{pradeep2022moby}, and Rangzen~\cite{lerner2016rangzen} all consider knowledgeable adversaries who optimally jam the densest areas.

\mypara{App-Specific Manipulations} Blackout-resistant apps may propagate additional information for improved routing and user experience; for example, Anix~\cite{kamali2024anix}, Moby~\cite{pradeep2022moby}, and Rangzen~\cite{lerner2016rangzen} each incorporate trust metrics to determine how messages spread and how they are presented to users.
Similarly, each work simulates how effectively adversaries can subvert the trust mechanism.
Anix supports a voting system for messages which is reinforced by users' social graphs; the paper then considers adversaries who spread misinformation while gaining the trust of benign users.
Similarly, Rangzen considers adversarial coalitions who attempt to bootstrap propaganda spread within their group before reaching other users.
Moby simulates adversaries who compromise trusted users to spread their messages.

\vspace{1mm} \hspace{-5mm}
\fbox{\parbox {0.97 \linewidth} {\textbf{Design goal 2.} \CTTF must consider and mitigate attacks that may be launched by non-global adversaries, including but not limited to denial-of-service attacks, network jamming, and application-specific manipulations.}}

\subsection{Evaluation Testbeds}
\label{subsec:evaluation-testbeds}
We now outline the testbeds and data used in each app's evaluation.
There is no suitable performance metric for comparing each work since they target different (but potentially overlapping) goals and functionalities.
Nevertheless, we highlight works that use real-world data versus those which use synthetic mobility models and comment on the trade-offs of each approach.
We also discuss the geographical scale of simulations and whether these can accurately reflect user actions in realistic blackout scenarios.

\mypara{Real-World Data} Most works we surveyed do not use real world mobility data in their simulations; those which do often have significant limitations.
For example, 1am~\cite{liu2015performance} uses data collected from 291 participants in a university town; however, due to the limited usage of their prototype the authors instead evaluated their performance in a simulation where each device initiated a message every six minutes.
This usage rate is over 7\,000 times higher than what was observed in their collected data.
Rangzen~\cite{lerner2016rangzen} uses mobility traces from 500 taxicabs~\cite{piorkowski2009crawdad}.
Unfortunately, automobiles travel too quickly for Bluetooth connections to be established in passing, so this dataset does not accurately reflect the real-world setup of pedestrians with smartphones.
Similarly, Liu et al.~\cite{liu_cooperative_2011} evaluate their system with the UMassDieselNet dataset which consists of connection data for 30 buses equipped with 802.11b Access Points over one semester at UMass Amherst. Lastly, 
Moby~\cite{pradeep2022moby} also uses real-world data for its evaluation but models user connections based on cell-tower data from 2009, which no longer accurately represents the proportion and usage of smartphones in the modern day.
Mirage~\cite{mirage} improves on prior works by using the same YJMob100K dataset as \CTTF, with GPS snapshots at 30-minute intervals.

\mypara{Synthetic Mobility Models} The remainder of the works we surveyed use synthetic data for evaluation.
The simplest mobility model employed by ASMesh~\cite{bienstock2023asmesh} and Anix~\cite{kamali2024anix} consists of users constantly (and randomly) moving through discrete grid cells during day and night.
This model fails to capture characteristics of real-world movement, such as travel between rural and metropolitan areas and increased travel during peak commuting hours.
BTM~\cite{rajagopalan_cross-layer_2006}, ORION~\cite{klemm_special-purpose_2003}, and M2MShare~\cite{palazzi_social-aware_2012} model users using random waypoints: each user picks a random position and moves towards it, repeating this process indefinitely.
Perry et. al.~\cite{perry2022strong} focus on protest scenarios but model crowds by placing users in a grid without evaluating how movement may affect system performance.
Amigo~\cite{inyangson2024amigo} also focuses on protest scenarios but greatly improves simulation realism by modelling common types of human protest movement, e.g, marches, chains, gatherings, and blockades.

\mypara{City-Scale} Prior works vary with the geographical scale of simulations.
Of the blackout-resistant messaging apps, only Amigo~\cite{inyangson2024amigo} and Perry et. al.~\cite{perry2022strong} evaluate their designs on dense protest scenarios.
BTM~\cite{rajagopalan_cross-layer_2006} and ORION~\cite{klemm_special-purpose_2003} each simulate their MANETs operating only over a $1 \text{km}^2$ area.
Liu et. al.'s~\cite{liu_cooperative_2011} simulations only operate over the bus routes on the UMass campus.
Much of the remaining work on blackout-resistant messaging and MANET-based file sharing evaluates performance at a city-wide scale, which is more appropriate for modelling connectivity during widespread outages.
However, as previously mentioned, both the real-world data and synthetic mobility models used in prior work are generally unrealistic and do not appropriately reflect real urban mobility.

\vspace{1mm} \hspace{-6mm}
\fbox{\parbox {0.97\linewidth} {\textbf{Design goal 3.} \CTTF should be evaluated via simulations which incorporate open source real-world data that accurately represents human movement at a city-wide scale.}}
\vspace{1mm}

Guided by the above goals, we begin by outlining the threat model \CTTF is designed to withstand, followed by a discussion of its architecture and workflow.

\section{Threat Model}
\label{sec:tm}

We assume a state-level adversary who can direct internet and cellular service providers to interrupt service within a city-wide region, thereby inducing an internet blackout. Before a blackout, the adversary may employ conventional censorship mechanisms such as filtering~\cite{ChinaThorns}, throttling~\cite{xue2021throttling}, or blocking~\cite{hoang2021great,ChinaActiveProbing}.
We also assume that other means of internet access, such as satellite broadband~\cite{hammas2025evading}, are either non-operational~\cite{quadri2024starlinkafrica}, forbidden~\cite{rbc2025iranstarlink}, or censored~\cite{gent2024chinafirewallspace}, forcing the vast majority of citizens within the affected region to make use of short-range communication technologies.

\mypara{Adversary capabilities}
Similarly to previous work~\cite{lerner2016rangzen,kamali2024anix}, we assume the adversary can operate Sybil nodes.
These nodes may perform arguably legitimate, yet malicious, app-specific manipulations, such as distributing poorly-chosen webpage ratings to negatively impact \CTTF's caching mechanisms.
Adversary nodes can also respond to requests from legitimate users untruthfully: for example, by refusing to provide a page they have stored, or by providing a page with manipulated contents.
Finally, the adversary may also perform signal jamming within specified land areas~\cite{pradeep2022moby} to entirely disable \CTTF's communication in a given range.

\mypara{Adversary constraints} Since \CTTF users communicate exclusively via short-range communication, we reason that it would be unfeasible for an adversary to globally monitor and intercept all user data exchanges.
Most prior works follow the same passive interception assumptions (\S\ref{subsec:considered-attacks}).
Further, we assume that although an adversary may have more centralized computational power than the average user (e.g., server-grade hardware); individual adversary devices (e.g., smartphones used as Sybil nodes) possess roughly equivalent computational power to the average user device.

\mypara{Non-goals}
We assume \CTTF is deployed where free speech is limited but governments stop short of large-scale arrests for using alternative communication methods~\cite{lerner2016rangzen,xue2024bridging}; e.g., adversaries will not isolate users to persecute them.
Prior works also aim to protect user identities from deanonymization, but \CTTF does not maintain identities.
Akin to prior works, \CTTF primarily uses Bluetooth classic, which can enable tracking via expensive equipment (2\,000+ USD) ~\cite{ludant2021linking}.
Nevertheless, all Bluetooth classic communication leaks MAC addresses (including everyday uses like streaming music); this limitation is shared by all prior works using Bluetooth, including those explicitly striving to provide anonymity ~\cite{lerner2016rangzen,pradeep2022moby,kamali2024anix}.
\wpes{\CTTF also supports BLE communication (\S\ref{sec:rating-exchange}) which rotates MAC addresses at regular intervals, but has lower throughput than Bluetooth classic.}

\section{Cache to the Future (\CTTF)}

\begin{figure*}[t]
    \centering
    \includegraphics[width=\textwidth]{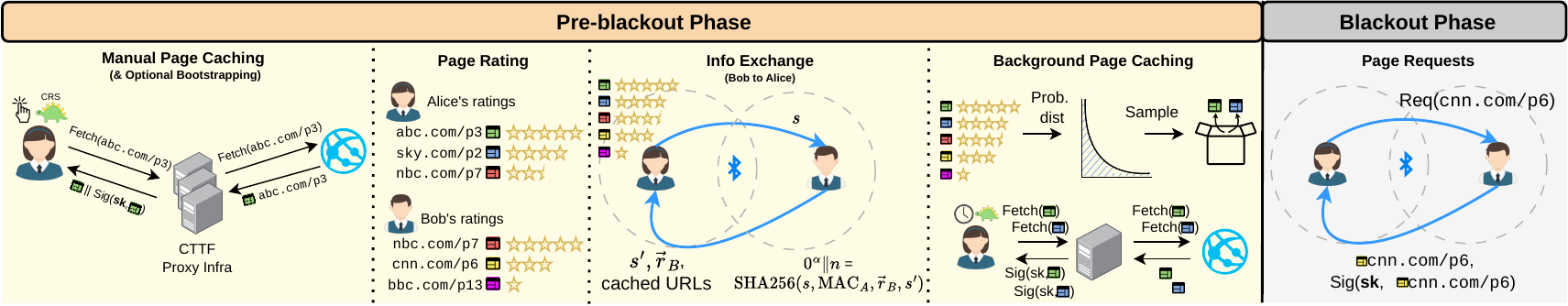}
    \vspace{-0.45cm}
    \caption{\CTTF's workflow. Users cache, rate, and exchange page ratings pre-blackout, requesting cached pages during blackout.}
    \label{fig:cttf-diagram}
    \vspace{-0.3cm}
\end{figure*}

This section addresses \CTTF's architectural elements and operational workflow, describing the system's core principles and functionality. Then, it elaborates on each of \CTTF's design aspects, before detailing its implementation.

\subsection{Architectural Overview}
\label{subsec:architecture}
\Cref{fig:cttf-diagram} depicts the architecture and operation for \CTTF.
The system is distributed over user devices, which are used in both the {\it pre-blackout} and {\it blackout} phases of \CTTF, and proxies, which are used only in the {\it pre-blackout} phase.
Users are within the adversary's geographical region of control whereas the proxies are assumed to be located in a censorship-free region.
User-to-user communication always occurs via short-range communication technologies while pre-blackout caching
can take place via HTTPS.

\subsection{\CTTF's Operational Workflow}
\CTTF relies on caching pages pre-blackout, exchanging ratings to determine which pages should be replicated, and then accessing the cached pages during a blackout.

\mypara{Cache bootstrapping}
When a user (say, Alice) browses a useful page, she can choose to cache it for future access in a blackout.
\CTTF will not fetch the page directly from the server for caching; instead, it fetches the page from a suite of trusted proxies which sign the page contents.
The signature allows other users to verify that Alice did not tamper with the page contents before distributing the page during the blackout.
Further details on the proxy service are discussed in \S\ref{sec:page-fetching}.
After caching pages, Alice can rate them on a scale of one to ten for perceived usefulness in a blackout, impacting the replication of these pages across other devices.

\mypara{Info exchange} Some users may install \CTTF pre-blackout while opting not to invest personal time into caching or rating pages.
Such devices can provide additional storage space for useful pages to be replicated across.
To determine which pages are useful, \CTTF locally exchanges page ratings and cached URLs when users come in close contact.
The rating exchange allows devices to compute an average usefulness value for each page.
We elaborate on the specifics of the rating exchange protocol and the method by which it limits adversarial influence in \S\ref{sec:rating-exchange}.

\mypara{Background caching}
In addition to pages being cached via manual selection, \CTTF will periodically cache pages in the background for users with additional storage space.
The background caching is performed proportionally to a user's locally computed page ratings.
This procedure involves transforming the page ratings into a probability distribution; the device then samples from this probability distribution to determine which pages to cache.
Caching then occurs via the same proxy previously described.
Further details on the probability distribution transformation are discussed in \S\ref{subsec:page-caching}.

\mypara{Page retrieval} During the blackout, users can view pages which they personally cached and pages their device cached in the background.
If a user wishes to view a page which has not been cached on their device, they can submit a request in the app to view said page.
This request is stored on the user's device until the page can be retrieved.
Pages are retrieved via local interaction with other \CTTF users; when in close enough proximity, a protocol runs in the background of each user's device which shares and retrieves requested pages over Bluetooth.
This process is detailed in \S\ref{sec:page-retrieval}.

\subsection{Page Fetching}
\label{sec:page-fetching}

When browsing webpages normally, the HTTPS protocol guarantees the integrity and authenticity of webpage data.
Contrastingly, in a blackout scenario, users of \CTTF fetch pages from each other.
As such, there is the concern that an adversary may tamper with the contents of a cached page to spread misinformation: for example, an adversary may modify content about \esoric{locations of abortion services or information about activism}.
To address this potential attack, \CTTF caches pages via trusted proxies which are assumed to be part of its trusted infrastructure (\S\ref{subsec:architecture}).

\mypara{Fetching via proxies} Rather than fetching a page's contents directly from the web, \CTTF will hit a proxy \wpes{from a suite of proxies} hosted outside of the censored region.
The contacted proxy is responsible for fetching the page on behalf of the user. 
The proxy also parses the retrieved HTML to fetch additional static resources (CSS, JS, images, etc.).
In addition, the proxy leverages digital signatures to guarantee the authenticity of the content during the blackout.
The input to the signing algorithm consists of the page contents, resources, and the time of retrieval.
The proxy returns to \CTTF the page contents as well as the associated signature.
\wpes{Valid public keys for proxies are} distributed with the \CTTF app, which allows users to verify the authenticity of pages received from other users.

\mypara{Reaching blocked proxies} Distributing information \wpes{on proxies} within the \CTTF app pre-blackout would increase susceptibility to censorship: an adversary could install the app and issue IP-based bans on any proxy addresses \CTTF attempts to connect to.
In general, we consider the details of how users find and connect to these proxies out-of-scope for \CTTF, but rely on existing censorship circumvention systems (CRS) to establish user-to-proxy connectivity.
For example, legitimate users could send an email/instant message to a proxy distribution service~\cite{tor_bridges} or connect with a proxy broker via domain fronting~\cite{fifield2015blocking}, similarly to how Tor users find bridges. A system like SpotProxy~\cite{kon2024spotproxy} could be utilized to rapidly refresh proxy IP addresses, thereby limiting a censor's ability to enumerate and block proxy IPs. 
Fully encrypted protocols~\cite{fenske2024bytes} such as Lyrebird~\cite{lyrebird} and Shadowsocks~\cite{shadowsocks} can also be leveraged to avoid detection and blocking of proxies by adversaries.

\mypara{Proxy-less operation} Should a \CTTF user be prevented from accessing \wpes{all trusted proxies from the suite} (e.g., due to internet censorship or transient outages), \CTTF can fallback to directly caching non-censored webpages without a proxy.
However, no trusted signature will be applied to the cached webpage.
Thus, while the page's contents can still be distributed, the cached information cannot be cryptographically verified during a blackout.
Nevertheless, \CTTF is compatible with different fallback mechanisms (\S\ref{sec:page-retrieval}) to assist the verification of page authenticity in such conditions.

\subsection{Info Exchange}
\label{sec:rating-exchange}

To feed \CTTF's automatic caching mechanism, a user's local \CTTF instance must first acquire page ratings from other users (i.e., community ratings), and then compute local averages to determine which pages are likely to be more valuable to cache. Below, we describe how \CTTF users establish short-range sessions using their smartphones and exchange these ratings. During this exposition, we also discuss mechanisms aimed at preventing adversarial nodes from significantly influencing community ratings.

\mypara{Session establishment}
\CTTF uses Bluetooth Low Energy (BLE) to establish sessions.
User devices broadcast BLE beacons indicating that they use \CTTF, \wpes{with two bits allocated in the broadcast to indicate which Bluetooth mediums they support (classic connections and/or BLE connections)}.
\wpes{Users who fear adversaries harvesting static MAC addresses can communicate exclusively over BLE connection-oriented channels (CoC) with short-term randomized MAC addresses at the cost of throughput}.
\CTTF continually scans for other \CTTF-compatible devices in the background.
\wpes{Upon discovering another device, a BLE CoC is opened.
If both devices support Bluetooth classic, they exchange their Bluetooth classic MAC addresses over BLE.
If Bluetooth classic MAC addresses were exchanged, the two devices can connect over Bluetooth classic in the background without needing to pair, otherwise, devices use the existing CoC.}
After a connection has been established, the rating exchange protocol takes place.

\mypara{Updating community ratings}
Each user transfers their ratings to the other during the session.
Their devices maintain a local average rating $a_i$ for each page rating $r_i$ they receive from other users, as well as the total number of ratings for each page ($c_i$).
$a_i$ is updated for each new page rating $r_i$ that is received.
$$a_i := (a_i \cdot c_i + r_i)/(c_i+1) \qquad c_i:= c_i+1$$
While \CTTF does not maintain identities, devices track recent MAC addresses to prevent legitimate users from over-contributing to local averages.
Because ratings guide \CTTF's background caching, adversaries may submit low scores for useful pages and high scores for irrelevant or deceptive pages. 
To limit adversarial influence, sessions are dropped if the adversary submits ratings outside the one to ten scale, or if the adversary submits a suspiciously high number of ratings.

\mypara{Thwarting rating manipulation}
Beyond malicious ratings, an adversary could bias local averages by spoofing their MAC address and reinitiating rating exchanges.
To mitigate this behaviour, users must compute a proof-of-work to share ratings.
To share his ratings vector $\vec{r}_B$, Bob receives a bitstring $s$ from Alice.
Bob must compute a bitstring $s'$ such that $\text{SHA256}(s, \text{MAC}_A, \vec{r}_B, s')$ starts with $\alpha$ 0s.
Bob returns $s'$ to Alice, who then recomputes the hash to verify the prefix before updating her local ratings.
We evaluate adversarial cache manipulation in 
\S\ref{subsec:eval-caching} and provide a concrete $\alpha$ in \S\ref{subsec:evaluation-microbenchmarks}. 
The entropy in $s$ ensures that PoWs cannot be precomputed and limits an adversary's ability to manipulate ratings; rapidly masquerading as new users incurs a high computational effort.

\mypara{Benefits of computing average ratings locally} \CTTF accumulates local community ratings fully locally; there is no centralized service collecting average user ratings, nor is there any routing protocol used to distribute these ratings.
A centralized rating service would be more highly susceptible to adversarial influence; even with a proof-of-work, an adversary may have access to significant computational resources that would outperform users submitting legitimate ratings.
Similarly, if users spread ratings on behalf of others via epidemic routing, an adversary could pre-compute proofs-of-work for large quantities of ratings from spoofed users.
Without user identities or any trust mechanism to restrict an adversary's ability to spread malicious ratings, any non-local strategy of averaging community ratings would be easily manipulable.

\mypara{Exchanging cached URLs} During rating exchange, each user also exchanges their set of cached URLs.
Over time, users accumulate a set of URLs cached by others which they can fetch during the blackout.
For increased usability, this set can be sorted according to each page's average rating.

\subsection{Caching Prioritization From Ratings}
\label{subsec:page-caching}
Users can cache pages through \CTTF manually at any point, e.g., during the initial bootstrap of the application and at any point prior to a blackout. However, \CTTF also uses any additional storage space granted to the application to automatically cache highly sought-after pages, according to community ratings received from other users.
Background caching occurs at periodical intervals (e.g., once a day) to ensure that cached pages are up-to-date when the blackout occurs. Below, we discuss potential caching strategies for \CTTF's automatic caching, and detail our solution.

\mypara{Caching strategies} There are a few potential strategies to choose which pages are cached based on a user's community ratings, each with pros and cons.
One potential strategy is to always cache the top $k$ rated pages, but this fails to replicate any more ``obscure'' content which may still prove useful to a small fraction of \CTTF users. As an alternative, one could linearly normalize ratings into a probability distribution and sample $k$ entries from this probability distribution; however, this approach may over-represent ``obscure'' content when sampling.
While we are not aware of any work which correlates user-driven webpage ratings with page popularity (nor would it be clear whether this correlation would be upheld in a blackout), it is relatively well-established that page access trends tend to follow a Zipf-like distribution~\cite{749260,lopes2024flow}.

\mypara{Caching via Zipf distribution} In \CTTF, we assume that page ratings follow a Zipf-like distribution, decreasing with the fraction of total accesses.
While no concrete data for page ratings is readily available, it is clear the transformation from ratings to a Zipf-like distribution requires some parametrization: it is unlikely for a user to give the most useful page a score of ten, the second most useful page a score of five, the third most useful page a score of three, etc., as a pure Zipf distribution would suggest.
We assume that ratings correlate with a transformed Zipf-like distribution:
$$f(x) = 10(a \cdot \text{Zipf}(x)^{b} + c)$$

The inverse of this transformation is used during sampling to convert ratings into probabilities.
\CTTF normalizes transformed ratings into a probability distribution and samples without replacement to determine which pages to cache.
We discuss appropriate constants $a$, $b$, and $c$ in \S\ref{subsec:page-rating-assumptions}.

\subsection{Page Retrieval}
\label{sec:page-retrieval}

During a blackout, users can no longer cache new pages, but they can instead request pages that may have been cached by other users.
While ``seeders'' may have access to the majority of pages which are of interest to them (e.g., because they manually cached them), the majority of ``leecher'' users depend on: a)  pages cached in the background based on cumulative community ratings exchanged pre-blackout, and; b) pages they request from other users. Below, we describe how page requests are fulfilled by \CTTF and how users may validate the received pages' authenticity. 

\mypara{Performing a request} A \CTTF user requests a page from her set of peer-cached URLs or by directly providing the URL she wishes to visit to the app. If the page has been cached in the device, the user is prompted: she can either view it immediately, or, alternatively, attempt the retrieval of a more recent version from other users. In this case, the request will be stored for sending in future encounters.

When two \CTTF users are nearby, they establish a device-to-device session, following the same process used to exchange page ratings (\S\ref{sec:rating-exchange}). Then, Alice and Bob execute the following page retrieval protocol. First, each user transmits a list of requested pages to the other.
If the other user stores a cached copy of the desired page, they will transmit this page and its original cached timestamp over Bluetooth; if the page is signed, they also transmit the corresponding proxy signature so that the page's authenticity can be verified (\S\ref{sec:page-fetching}).
The user can view the page on their device and mark the request as fulfilled after it has been successfully retrieved.
Any unfulfilled requests persist on user devices until at least the next encounter.

\mypara{Validating received pages} If Alice receives a page from Bob with a signature, she verifies its authenticity; otherwise, the request is kept on Alice's device.
Future encounters may yield a signed copy or allow a fallback authenticity check.
For instance, Alice can take a majority vote, collecting page copies from different users and selecting the most common version; \wpes{we evaluate this fallback in \S\ref{subsec:comparison-models}}.
Alice could also compute perceptual hashes~\cite{yangcertphash} across multiple versions of the page, discarding outlier variants \wpes{and identifying high-similarity variants as trusted} or \wpes{meet with friends and family members to retrieve a page from a trusted contact}.
We relegate the exploration of alternative validation mechanisms to future work.%

\mypara{Freshness resolution}
If Alice requests an update for a page she previously cached and later receives an additional copy, she compares her page's timestamp with the received timestamp.
If the newly received page has an older timestamp (bounded by some interval, e.g., 24h), this data is discarded.
Otherwise, if the received page can be validated (either by a signature or a fallback mechanism), it replaces Alice's cached page.
When Alice's cached page is replaced, she can choose to keep the page in her request list, aiming to find an even more recent version, or mark her request as fulfilled.

\section{Evaluation Methodology}

This section describes our evaluation methods; results are later presented in \S\ref{sec:evaluation-results}.
We begin by discussing goals and metrics in \S\ref{subsec:evaluation-goals}.
From there, we discuss the city-scale mobility data used for modelling in \S\ref{subsec:mobility-models}.
Finally, in \S\ref{subsec:simulator-param-configs} we detail the simulator we developed and the parameters we customize in our evaluation.

\subsection{Evaluation Goals and Metrics}
\label{subsec:evaluation-goals}
We evaluate \CTTF in benign and hostile blackouts via:

\mypara{Page request satisfaction} We evaluate request completion: the fraction of requests eventually resolved, across usefulness ratings.
We also measure retrieval latency: the expected duration for a requested page to be retrieved.
Finally, we assess how adversarial behaviour, such as Sybil nodes performing DoS attacks and using signal jamming to prevent communication, affects these metrics.

\mypara{Robustness to cache manipulation} \CTTF relies on replicating pages across devices according to user ratings. We wish to assess how effectively \CTTF's local averaging and proof-of-work mechanisms resist adversarial manipulation.
We measure the proportion of users which cache pages at the head and tail of the distribution as we vary the fraction of adversarial Sybil nodes in the network.

\mypara{Comparative approaches} We validate \CTTF's data dissemination approach: that using ``leecher'' devices to replicate pages is more effective than retrieving pages from ``seeders'' via epidemic routing.
We present a comparative simulation and evaluate it according to the metrics of satisfaction and average latency.  
We also evaluate \CTTF using alternative mobility datasets sourced from prior works.

\mypara{Application performance} Finally, we evaluate \CTTF's resource burden through micro-benchmarks on a prototype app.
We measure throughput, battery consumption, and the time required to complete PoWs at various difficulties.

\subsection{City-wide User Mobility Model}
\label{subsec:mobility-models}

\wpes{Our simulations aim to show \CTTF's efficacy in a realistic city-scale environment.
To this end, we based our model on recent real-world mobility data sourced from the YJMob100K dataset~\cite{yabe2024yjmob100k}. For easing comparisons to existing work, we also assess \CTTF's performance in the idealized grid model used by ASMesh~\cite{bienstock2023asmesh} and Anix~\cite{kamali2024anix}.}

\mypara{YJMob100K} The YJMob100K dataset contains 75 days of mobility data for 25\,000 individuals living within a highly-populated Japanese city updated at 30 minute time intervals.
The city is discretized into a $200\times200$ grid of 500 meter by 500 meter cells.
The 25\,000 individuals make up 1.25\% of the city's population.
Point-of-interest annotations for cells allow us to distinguish between metropolitan and rural areas.
\esoric{Since cells may exceed Bluetooth range, our evaluation includes experiments run with {\it reduced contact probability} to model infrequent interactions (\S\ref{subsec:main-results}).}

\begin{figure}[t]
    \centering
    \includegraphics[width=0.7\linewidth]{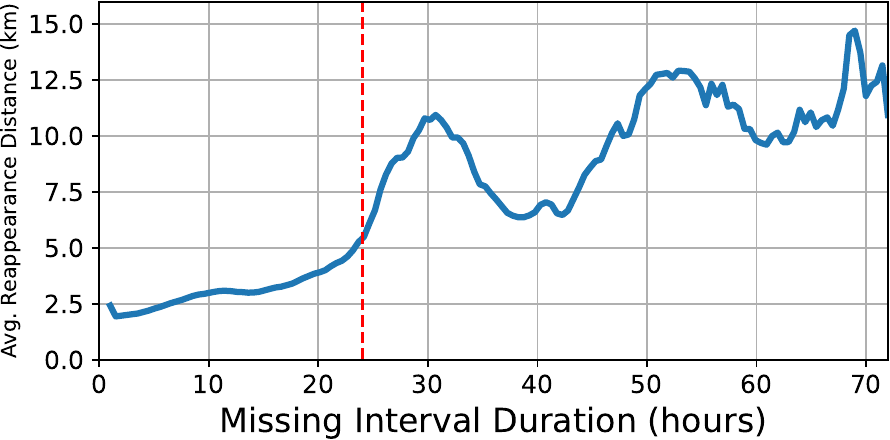}
    \vspace{-0.2cm}
    \caption{Average reappearance distance vs. missing interval duration. The red dashed line marks the 24-hour cutoff for applying linear interpolation.}
    \label{fig:interpolation}
    \vspace{-0.5cm}
\end{figure}

\myparait{Mobility data interpolation} YJMob100K sampled less frequently at low movement speeds to limit battery drain.
We fill in missing data via linear interpolation.
99\% of gaps were under 24h, with users reappearing within $3\text{km}$ on average; reappearance distance increased substantially for intervals greater than 24h (see \Cref{fig:interpolation}).
For sub-24h gaps, we interpolate positions between the grid cells immediately before and after.

\mypara{Random movement grid} \wpes{As discussed in \S\ref{subsec:evaluation-testbeds}, prior works have been evaluated with a large variety of datasets and testbeds which makes direct comparison challenging.
Nevertheless, as a best-effort comparison, we implemented the random grid-based movement used by select prior works~\cite{kamali2024anix,bienstock2023asmesh} in our simulator.
In this model, users are placed in an $N \times N$ grid and move randomly to a nearby tile at each timestep; users within the same grid cell \cite{kamali2024anix} or surrounding grid cells \cite{bienstock2023asmesh} at a given timestep are assumed to be within communication distance.
In our evaluation, we show that \CTTF is also effective in this model (\S\ref{subsec:comparison-models}).}

\subsection{Simulator and Parameter Configurations}
\label{subsec:simulator-param-configs}

We now discuss \CTTF's simulator, listing parameters, their purposes, and selected values rationale.
For a summary, see \Cref{table:parameters} (App.~\ref{appendix:simulation-parameters}).

\mypara{Experimental testbed} The simulator comprises 1\,300 lines of Python using \texttt{scipy}, \texttt{numpy}, and \texttt{pandas}.
Experiments ran in parallel on a machine with 8 Intel Xeon Platinum 8276 CPUs and 6TB of RAM.
Before the blackout, users interact and exchange ratings, after which pages are sampled and cached.
During the blackout, users request pages and interact with others to obtain them.

\mypara{Users' distribution} Our simulation supports three user types.
\textit{Leechers} install \CTTF pre-blackout but do not manually cache or rate pages, simply accumulating ratings from others and caching automatically.
\textit{Seeders} actively cache and rate pages based on perceived usefulness.
\textit{Adversaries} disrupt the system via rating manipulation, DoS, and jamming.
All users follow a unique individual's mobility trace in the YJMob100K dataset.
Our baseline experiments assume 2\% adversaries~\cite{lerner2016rangzen} with remaining split 75\%/25\% leechers/seeders.
This is a conservative estimate: prior BitTorrent studies found seeders typically outnumber leechers~\cite{meulpolder2010public,cox2010seeders}.
In a benign scenario, varying the leecher to seeder ratio did not significantly impact results (App.~\ref{appendix:additional-eval}).

\begin{figure}[t]
    \centering
\includegraphics[width=0.30\textwidth]{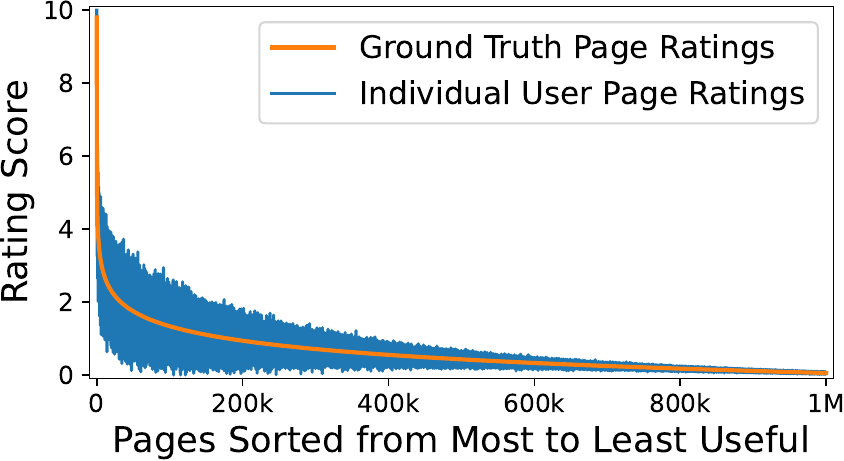}
    \vspace{-0.2cm}
    \caption{Ground truth and simulations' user page ratings.}
    \label{fig:page-ratings}
    \vspace{-0.5cm}
\end{figure}

\mypara{Page ratings and exchange} The page universe in our simulation consists of one million pages; this value was chosen since the top million sites on the Internet capture over 95\% of all page loads and time spent online~\cite{ruth2022world}.
Each user caches 1\,200 pages and associated static content.
With a median mobile page size of 2.5 MB~\cite{HTTPArchive} this corresponds to approximately 3 GB of user device space allocated to \CTTF.
In our baseline, we assume each seeder rates 500 pages; increasing seeder ratings did not significantly impact results (App.~\ref{appendix:additional-eval}).
Since the average web user visits approximately 150 pages per day~\cite{crichton2021home}, seeders could rate this many pages given a few days of participation before the blackout.
To limit adversarial influence, users can exchange a maximum of 1\,000 page ratings in any individual interaction.

\mypara{Page rating assumptions} \label{subsec:page-rating-assumptions}
We assume a ground-truth Zipf-like distribution for page accesses which is transformed into a similarly shaped curve for user ratings.
A diagram is shown in \Cref{fig:page-ratings}.
Each user's ratings follow the global ground-truth ratings with additive exponentially decaying normal noise.
The constants for the rating to probability distribution described in \S\ref{subsec:page-caching} were chosen to align with a recent internet use study; Ruth et al.~\cite{ruth2022world} found that the top 10, 100, and 10\,000 pages receive 25\%, 40\%, and 70\% of all hits respectively, so we assign ratings 10-7.5, 7.5-6, and 6-3 for pages 1-10, 11-100, and 100-10\,000.
We fit $f$ via \texttt{scipy} optimization with constraints $f(1) = 10, f(10) = 7.5, f(100) = 6, f(10\,000) = 3$ and $f(1\,000\,000) = 0$, finding that $a = 2.34$, $b = 0.039$, and $c = -1.36$ minimized squared error.

\mypara{Page requests}
Users request pages with 25\% probability at each blackout timestep (i.e., averaging one request every 2 hours).
When colocated, users interact for a 1 minute duration with a configurable contact probability.
\esoric{Due to city density} we assume users interact with 100\% probability but also evaluate sparse scenarios with infrequent contact in \S\ref{subsec:main-results}.
Given Bluetooth classic throughput (\S\ref{subsec:evaluation-microbenchmarks}) and a 2.5MB median page size, each user transmits at most four pages per session.

\mypara{Simulation grid parameters}
\wpes{The YJMob100K dataset has a grid size of $200 \times 200$ with $25\,000$ users, however, for our random grid movement experiments, we use the same $25 \times 25$ grid with $600$ users as configured by prior works~\cite{kamali2024anix,bienstock2023asmesh}.
In the YJMob100K experiments, users' movement is dictated by the data included in the dataset; for the random grid experiments, we assume users can move a maximum of 2 grid cells in any direction, per timestep.}

\begin{figure*}[ht!]
  \centering

  \subfloat[Jamming 0 $\text{km}^2$]{
    \includegraphics[width=0.14\textwidth]{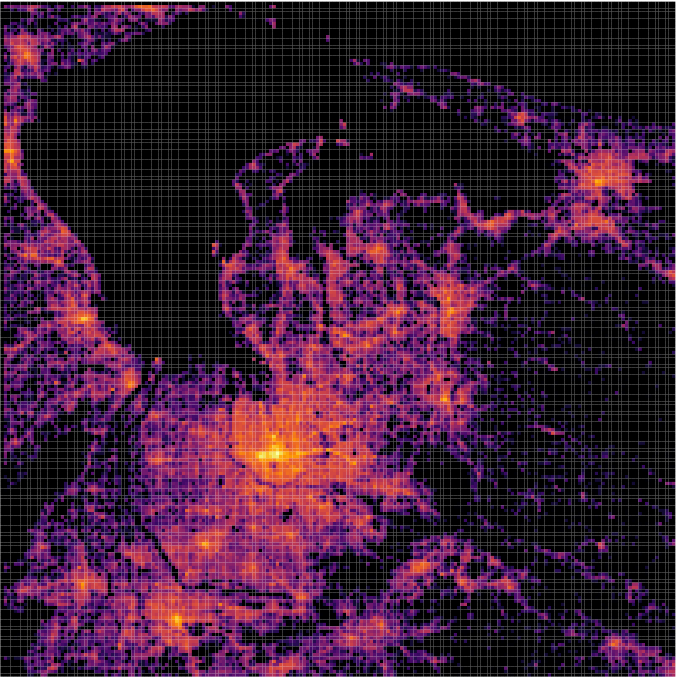}
  }
  \hspace{0.25cm}
  \subfloat[Jamming 2.5 $\text{km}^2$]{
    \includegraphics[width=0.14\textwidth]{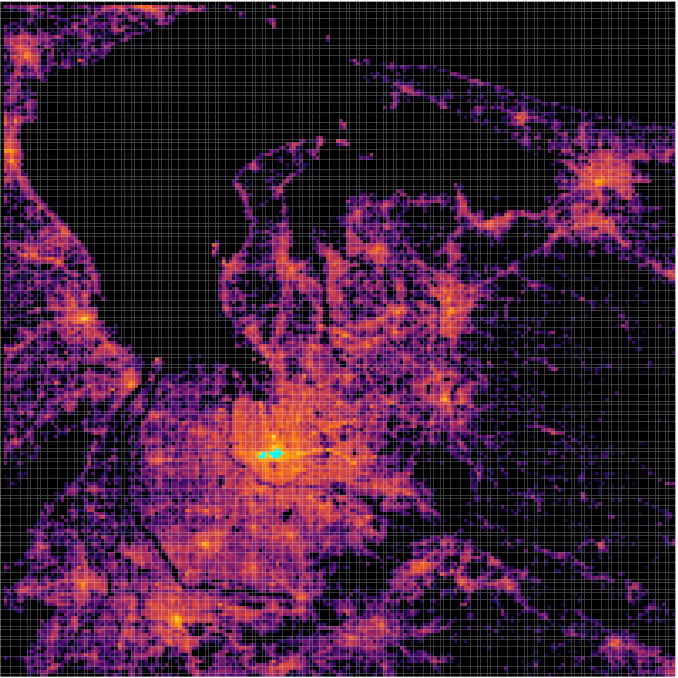}
  }
  \hspace{0.25cm}
  \subfloat[Jamming 25 $\text{km}^2$]{
    \includegraphics[width=0.14\textwidth]{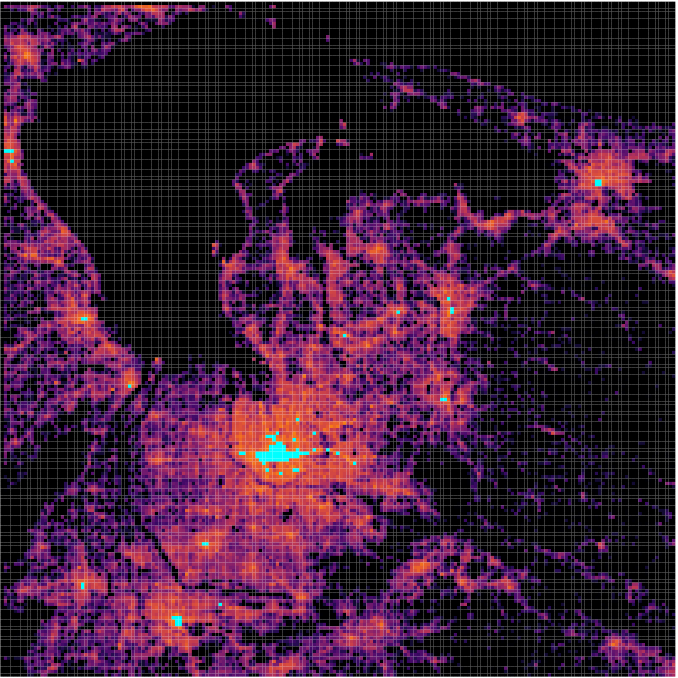}
  }
  \hspace{0.25cm}
  \subfloat[Jamming 250 $\text{km}^2$]{
    \includegraphics[width=0.14\textwidth]{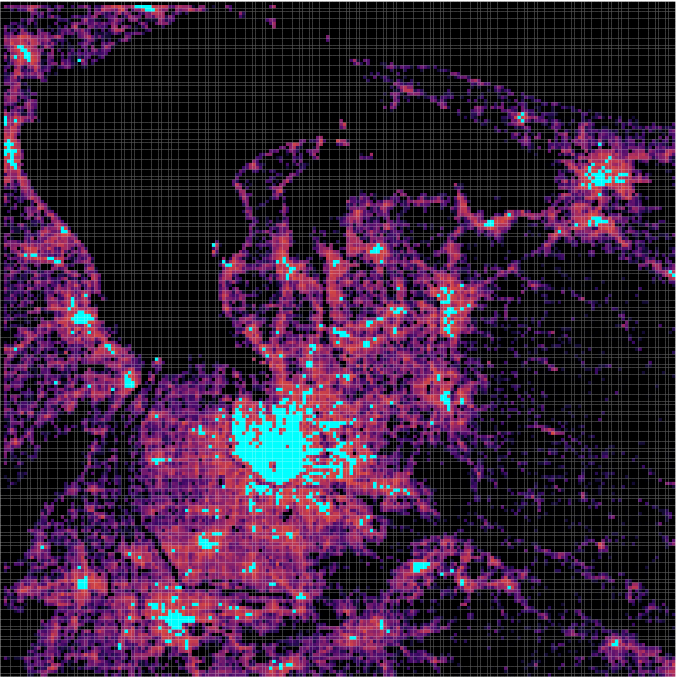}
  }
  \hspace{0.25cm}
  \subfloat[Jamming 2\,500 $\text{km}^2$]{
    \includegraphics[width=0.14\textwidth]{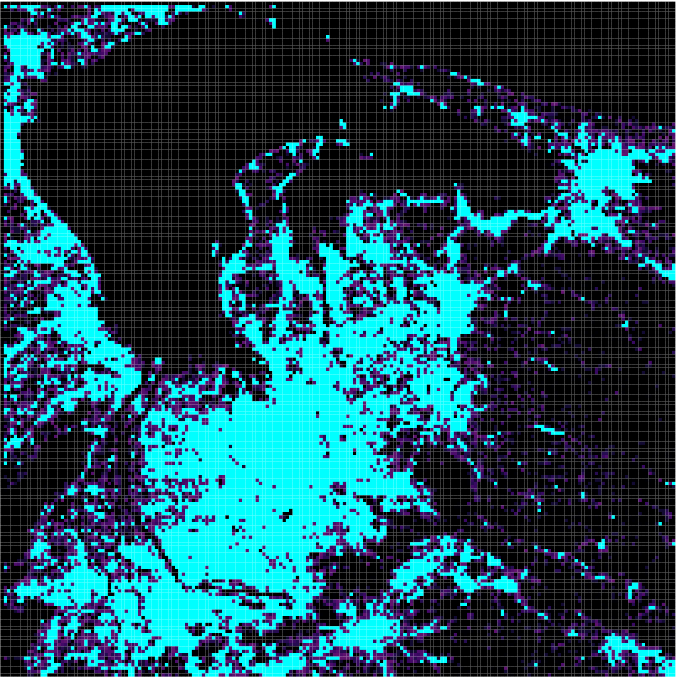}
  }
  \hspace{0.05cm}
  \subfloat{
    \includegraphics[width=0.052\textwidth]{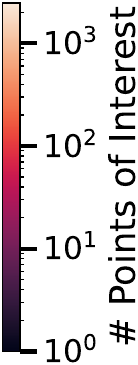}
    \vspace{0.4cm}
  }

  \caption{YJMob100K area with cyan cells jammed, prioritizing those with most points of interest.}
  \label{fig:jamming-heatmap}
  \vspace{-0.25cm}
\end{figure*}
\begin{figure*}[!t]
  \centering

  \subfloat[Request satisfaction\label{fig:benign-request-satisfaction}]{
    \includegraphics[width=0.28\textwidth]{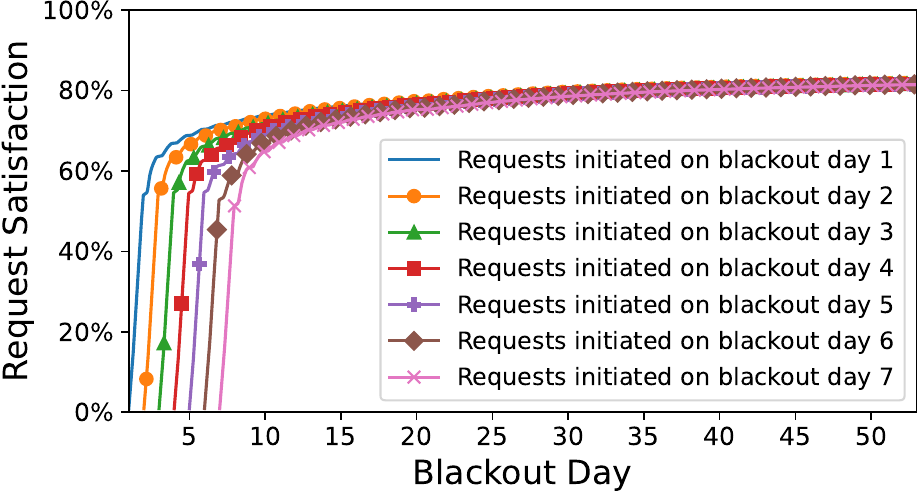}
  }
  \hfill
  \subfloat[Latency\label{fig:benign-box-whisker}]{
    \includegraphics[width=0.28\textwidth]{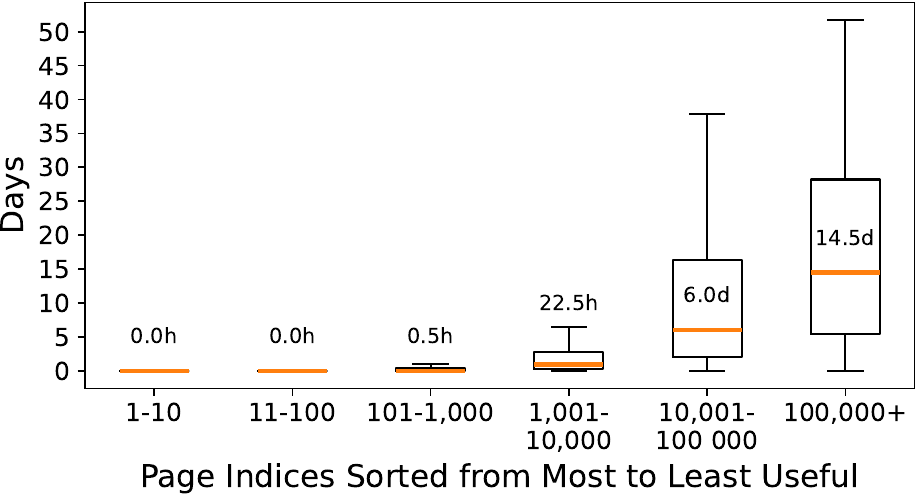}
  }
  \hfill
  \subfloat[Satisfaction by Index\label{fig:benign-index-satsifaction}]{
    \includegraphics[width=0.28\textwidth]{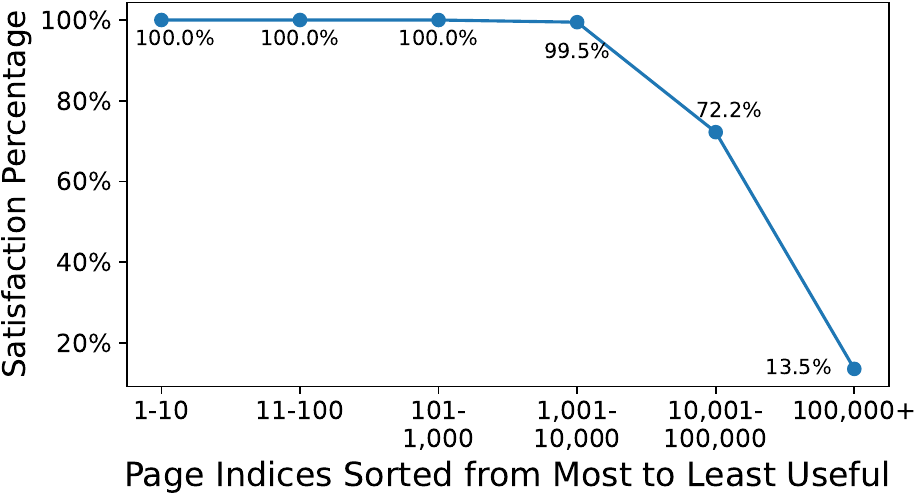}
  }
  \vspace{-0.3cm}
  \caption{Request satisfaction, latency to fulfill page requests, and request satisfaction by page index (\textit{benign} scenario).}
  \vspace{-0.25cm}
  \label{fig:benign}
\end{figure*}\begin{figure*}[!t]
  \centering

  \subfloat[Request satisfaction \label{fig:contact-request-satisfaction}]{
    \includegraphics[width=0.28\textwidth]{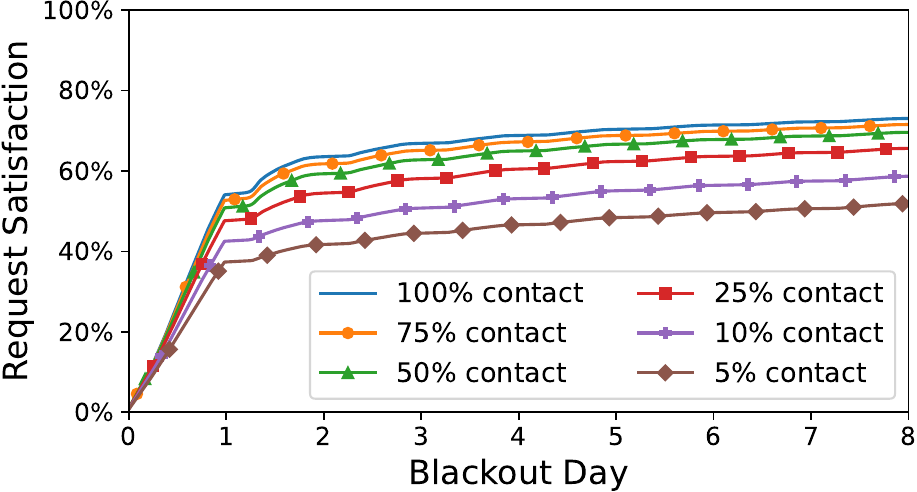}
  }
  \hfill
  \subfloat[Latency\label{fig:contact-box-whisker}]{
    \includegraphics[width=0.28\textwidth]{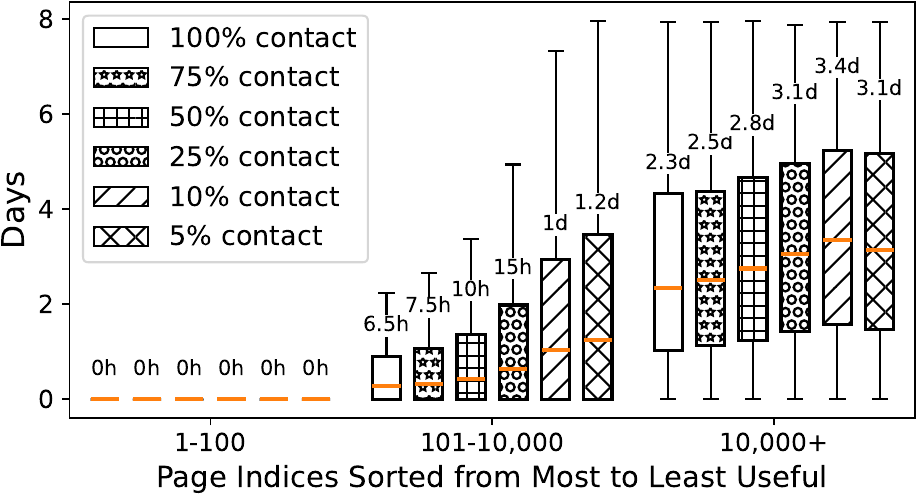}
  }
  \hfill
  \subfloat[Request satisfaction across indices \label{fig:contact-satisfaction-indices}]{
    \includegraphics[width=0.28\textwidth]{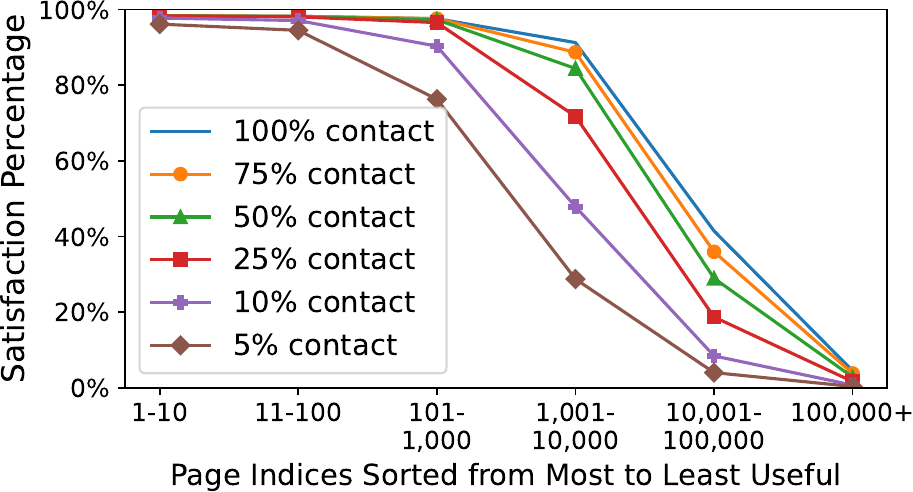}
  }
  \vspace{-0.3cm}
  \caption{Impact of contact probability (\textit{benign} scenario). Text above box plots shows median latency for satisfied requests.}
  \label{fig:contact}
  \vspace{-0.0cm}
\end{figure*}

\mypara{Adversarial behaviour}
All adversarial nodes cooperate by giving the top 500 most useful pages a score of 1 and the bottom 500 pages a score of 10, submitting ratings once per interaction.
We also evaluate the efficacy of proofs-of-work in a ``stalking scenario'' where each adversary tracks one leecher within their grid cell and repeatedly reinitiates rating exchange; we parameterize the proof-of-work at 1 minute, 30 seconds, and 1 second per exchange across different proportions of adversaries.
Finally, adversaries may signal jam grid cells; 
we assume rational adversaries, jamming the top $k$ grid cells with the most points of interest (i.e., highest population). Signal jamming heatmaps are displayed in \Cref{fig:jamming-heatmap}.

\mypara{Epidemic routing}
For our epidemic routing experiments, users store at most 300 pages for forwarding ($\sim$750 MB).
No ratings are exchanged and leechers cache nothing; all pages are fetched from seeders during the blackout.
Due to the 4-page session limit (see \textbf{Page requests.}), forwarding is severely limited.
Adversaries request and forward pages drawn uniformly randomly from the latter-half of the page universe (i.e., less useful pages), and forward responses similarly.

\section{Evaluation Results}
\label{sec:evaluation-results}

We now present our evaluation's results.
Simulations begin with a pre-blackout rating exchange, after which non-adversaries cache based on local ratings. The blackout phase has two parts: a page-requesting period, and a long-term observation phase tracking request satisfaction.
We evaluate latency and satisfaction for benign (\S\ref{subsec:main-results}) and adversarial (\S\ref{subsec:main-results-adversarial}) settings, adversarial caching influence (\S\ref{subsec:eval-caching}), model comparisons (\S\ref{subsec:comparison-models}), and microbenchmarks (\S\ref{subsec:evaluation-microbenchmarks}).

\subsection{\CTTF Performance in Benign Scenarios}
\label{subsec:main-results}

\mypara{\CTTF provides fast access to popular pages in a benign scenario} 
To observe \CTTF's performance, we configured a 60-day benign scenario simulation with one week of rating exchanges pre-blackout, and one week of page requests during the blackout.
The requesting period is capped to observe how satisfaction converges over a long time period.
Request satisfaction performance over time is presented in \Cref{fig:benign-request-satisfaction}.
Requests follow a similar curve no matter which day they were initiated on, reaching 54\% satisfaction after one day, 72\% satisfaction after one week, and 82\% satisfaction at the end of the two month simulation.
Box plots for system latency in this scenario are presented in \Cref{fig:benign-box-whisker}: pages with utility rating above 3 can be fetched within a day on average.
The top 1\,000 pages can be fetched in less than 30 minutes on average due to their high prevalence across devices.
\Cref{fig:benign-index-satsifaction} presents request satisfaction by page indices: requests for the 100\,000 most useful pages are satisfied with high probability; requests going unsatisfied mostly belong to the long tail of the page utility distribution.

\mypara{Follow-up simulations} Unlike our initial 2-month simulation, the remaining experiments have one week of rating exchanges, one day where non-adversaries request and receive pages, followed by one week with no additional page requests to observe the effect of user mobility on satisfaction.
Shorter simulations allowed testing a wider range of parameters within our computational constraints.

\mypara{Limited contact remains effective}
Figs. \labelcref{fig:contact-request-satisfaction}, \labelcref{fig:contact-box-whisker} and \labelcref{fig:contact-satisfaction-indices} present request satisfaction and latency as we adjust the probability that two users in the same cell interact.
Using a 5\% contact probability rather than a 100\% contact probability drops request satisfaction after 1 week to 52\% from 73\%.
\Cref{fig:contact-satisfaction-indices} indicates that this drop-off impacts lower utility pages more substantially.
Latency does not increase for the top 100 pages, increases from an average of 6.5 hours to 1.2 days for pages 101-10\,000, and from 2.3 days to 3.1 days for all remaining pages.
This suggests that \CTTF allows users to retrieve pages even assuming low contact between users in the same cell.
Nevertheless, as in prior works \cite{pradeep2022moby,kamali2024anix,bienstock2023asmesh}, we fix contact probability at 100\% while evaluating how other factors effect performance.

\subsection{\CTTF Performance in Adversarial Scenarios}
\label{subsec:main-results-adversarial}

\begin{figure*}[]
  \centering
  
    \subfloat[Request satisfaction\label{fig:varyAdversary-request-satisfaction}]{
    \includegraphics[width=0.28\textwidth]{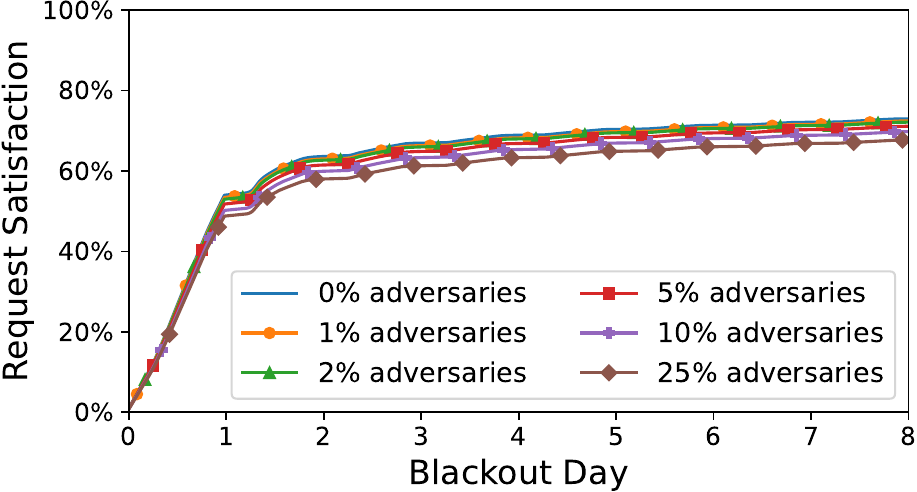}
  }
  \hfill
   \subfloat[Latency\label{fig:varyAdversary-box-whisker}]{
    \includegraphics[width=0.28\textwidth]{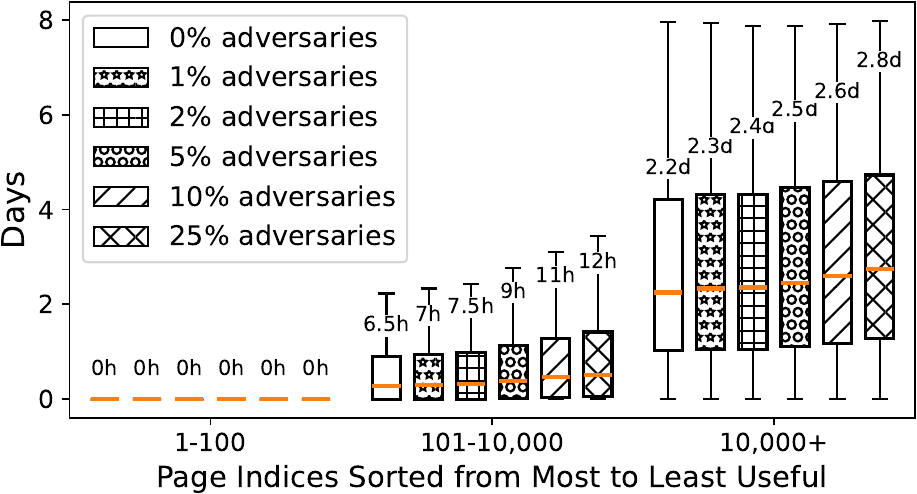}
  }
  \hfill
  \subfloat[Request satisfaction across indices \label{fig:varyAdversary-satisfaction-indices}]{
    \includegraphics[width=0.28\textwidth]{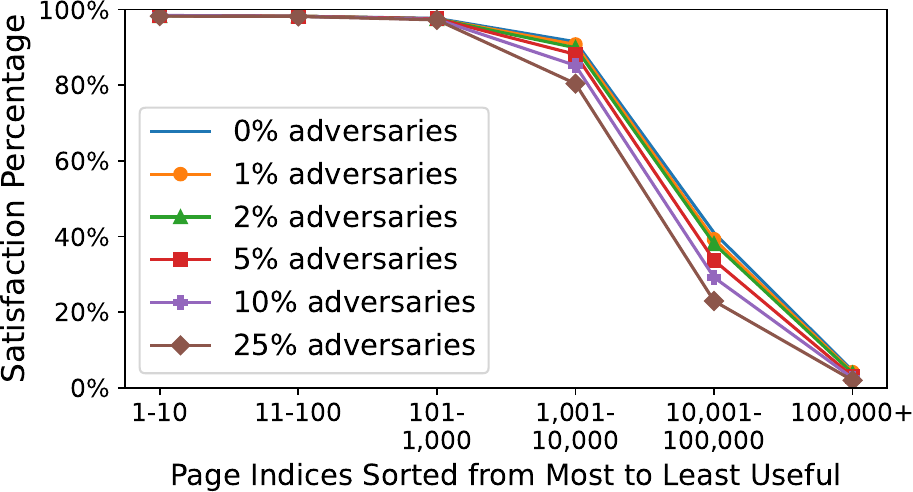}
  }
  \vspace{-0.3cm}
  \caption{Impact of Sybil nodes (\textit{adversarial} scenario).}
  \label{fig:varyAdversary}
\end{figure*} 

\mypara{Sybils cannot significantly disrupt \CTTF}
Figs. \labelcref{fig:varyAdversary-request-satisfaction}, \labelcref{fig:varyAdversary-box-whisker} and \labelcref{fig:varyAdversary-satisfaction-indices} present request satisfaction and latency as the Sybil percentage varies, with leechers and seeders comprising 75\% and 25\% of the remaining nodes, respectively.
Even as Sybil nodes increase to 25\% of the network, request satisfaction after one week drops only 5\% from a benign scenario and latency increases by 15 hours on average for low utility pages.
These results suggest that \CTTF's caching mechanism resists adversarial manipulation even with high proportions of Sybil nodes (further expounded upon in \S\ref{subsec:eval-caching}).

\begin{figure*}[]
  \centering
  
    \subfloat[Request satisfaction\label{fig:jamming-request-satisfaction}]{
    \includegraphics[width=0.28\textwidth]{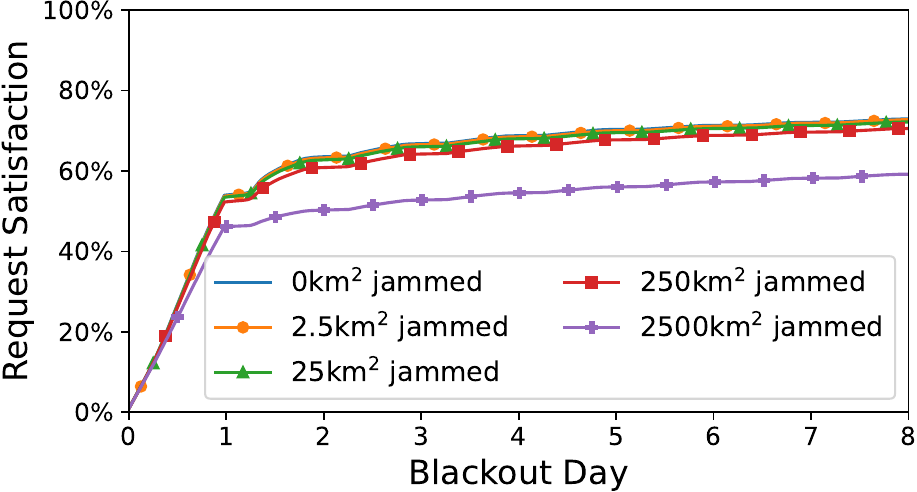}
  }
  \hfill
\subfloat[Latency\label{fig:jamming-box-whisker}]{
    \includegraphics[width=0.28\textwidth]{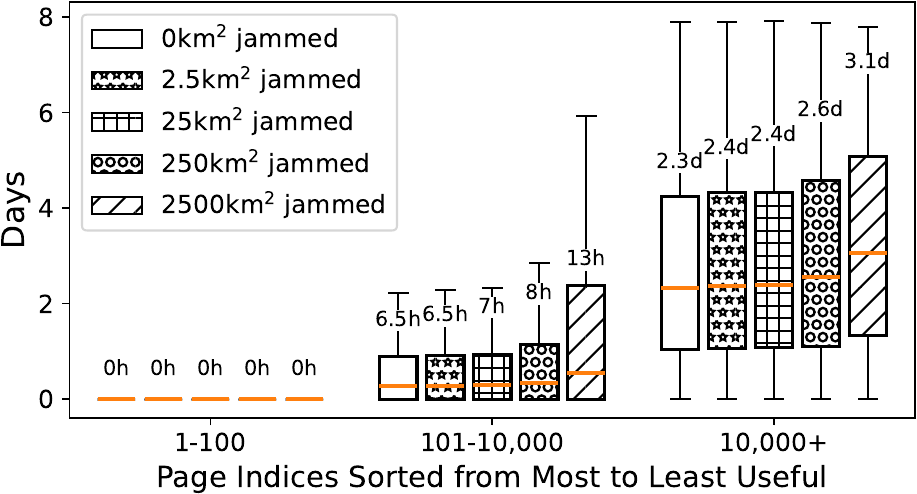}
  }
  \hfill
  \subfloat[Request satisfaction across indices\label{fig:jamming-satisfaction-indices}]{
    \includegraphics[width=0.28\textwidth]{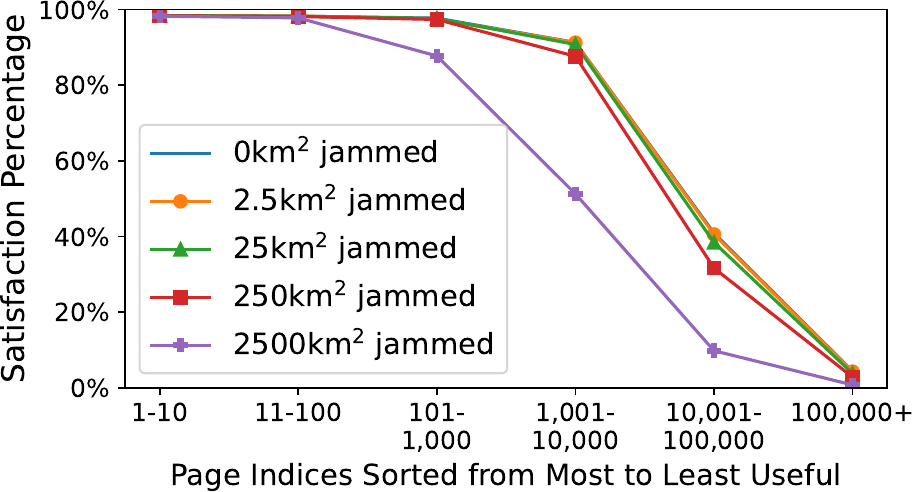}
  }
  \vspace{-0.3cm}
  \caption{Impact of jammed cells (\textit{adversarial} scenario).}
  \label{fig:jamming}
\end{figure*}

\mypara{Jamming must be broad to disrupt \CTTF}
Figs. \labelcref{fig:jamming-request-satisfaction}, \labelcref{fig:jamming-box-whisker} and \labelcref{fig:jamming-satisfaction-indices} present satisfaction and latency as we adjust the jammed area.
Jamming 250 $\text{km}^2$ of the most point-of-interest dense locations only reduces satisfaction by 2\% and only increases latency by a couple of hours on average.
An adversary would need to jam an unfeasibly large area of 2\,500 $\text{km}^2$ to significantly impact \CTTF.

\begin{figure}
  \subfloat[Req. satisfaction (after a week). \label{fig:jammingAndSybil-satisfaction}]{
    \includegraphics[width=0.45\linewidth]{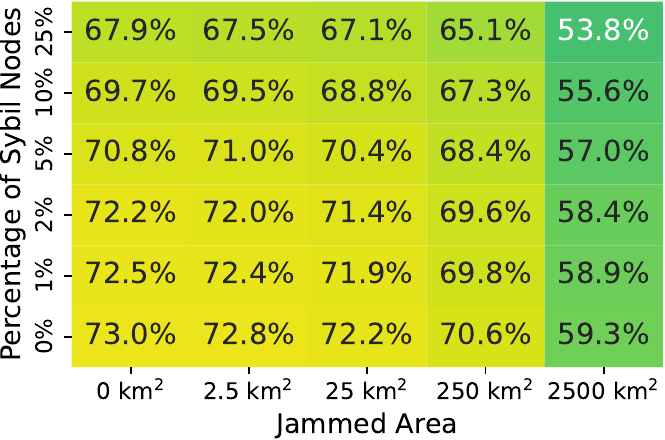}
  }
  \hfill
  \subfloat[90$^{th}$ percentile latency when requesting top-10k most useful pages. \label{fig:jammingAndSybil-latency}]{
    \includegraphics[width=0.45\linewidth]{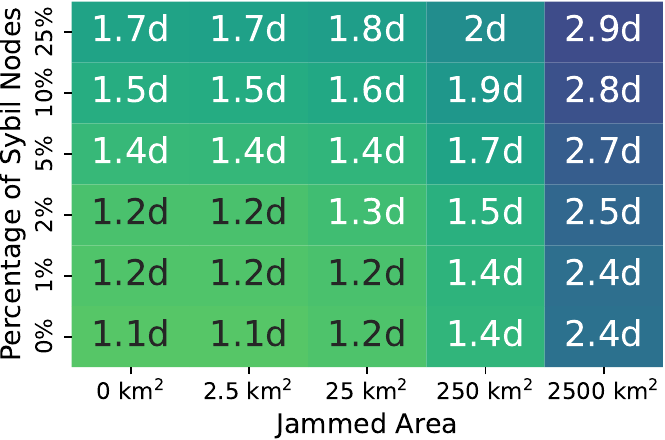}
    
  }    
  \caption{\CTTF metrics with jamming and Sybils.}
    \label{fig:enter-label}
\end{figure}

\mypara{Requests can still be satisfied w/ Sybils and jamming} 
Figs. \ref{fig:jammingAndSybil-satisfaction} and \ref{fig:jammingAndSybil-latency} present heatmaps for request satisfaction and latency when adversaries deploy Sybil nodes and jam cells.
Even under aggressive conditions such as 10\% Sybil nodes and 250 $\text{km}^2$ jammed, \CTTF can satisfy 67\% of all requests, with requests for the top-10\,000 pages having a 90\% chance of being satisfied in under two days.

\subsection{Caching with Adversarial Influence}
\label{subsec:eval-caching}
\begin{figure}[!t]
    \centering
    \includegraphics[width=0.7\linewidth]{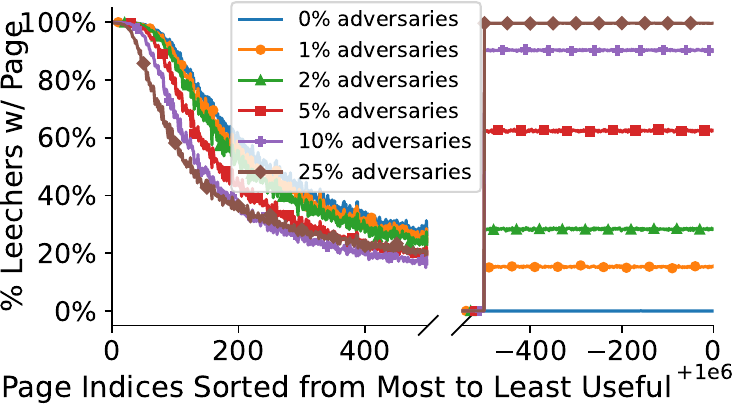}
    \vspace{-0.3cm}
    \caption{Leecher caching with Sybil nodes.}
    \label{fig:pages-cached-by-leechers}
\end{figure}

\mypara{Adversaries cannot prevent caching of popular pages}
In our simulation, all adversaries cooperate to manipulate ratings for the top 500 most useful and least useful pages.
Sufficiently high proportions of coordinated adversaries ($\geq 5\%$) are able to blunt ratings for useful pages and significantly increase ratings for less useful pages (App.~\ref{appendix:additional-eval} - Fig.~\ref{fig:average-leecher-ratings}).
The effect of the rating manipulation on the cache is shown in \Cref{fig:pages-cached-by-leechers}: at 2\% Sybil nodes, adversaries are able to convince 25\% of leechers to cache useless pages.
However, as illustrated in \Cref{fig:varyAdversary}, the cache manipulation has a non-prohibitive impact on request satisfaction and latency, \wpes{indicating that highly sought-after pages are still being served}. 

\begin{figure}[!t]
    \centering
    \includegraphics[width=0.7\linewidth]{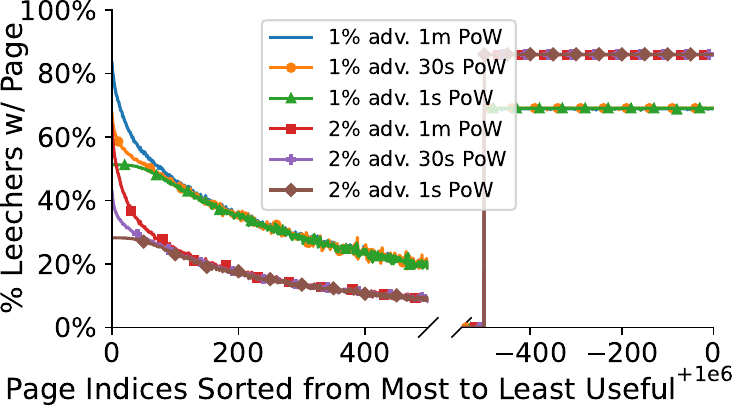}
    \vspace{-0.3cm}
    \caption{Leecher caching under stalking scenario.}
    \label{fig:pages-cached-by-leechers-stalking}
\end{figure}

\mypara{Proofs-of-work can blunt adversarial influence}
\Cref{fig:pages-cached-by-leechers-stalking} presents the effects of instituting proofs-of-work in a simulation where adversaries stalk leechers while rapidly spoofing their Bluetooth MAC address to re-initiate rating exchange.
This behaviour allows adversaries to more effectively reduce caching of popular pages; however, a 1 minute proof-of-work noticeably mitigates this effect for the top 100 pages.
In addition, the stalking scenario across each parametrization shown maintains an overall request satisfaction rate of 68\%, with a near 100\% request satisfaction rate for the first 1\,000 pages, indicating that seeders are still serving useful pages even if leecher ratings are successfully manipulated by adversaries.
This setup assumes adversary devices cannot circumvent the blackout or use cellular networks to forward the PoW to server-grade hardware.
Nevertheless, the 1s PoW lines indicate that leechers still cache popular pages, albeit with a worse probability, even if PoWs are easily cracked.
\wpes{Future work could further explore adversary spam mitigation by limiting interactions based on elapsed time, distance travelled, and local population density.}


\subsection{Comparisons Against Other Models}
\label{subsec:comparison-models}
\begin{figure}[t]
  \subfloat[Req. satisfaction (after a week). \label{fig:epidemic-satisfaction}]{
    \includegraphics[width=0.45\linewidth]{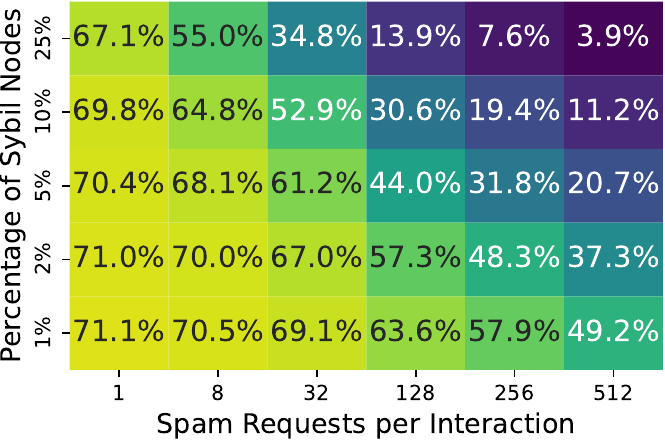}
  }
  \hfill
  \subfloat[90$^{th}$ percentile latency when requesting top-10k most useful pages. \label{fig:epidemic-latency}]{
    \includegraphics[width=0.45\linewidth]{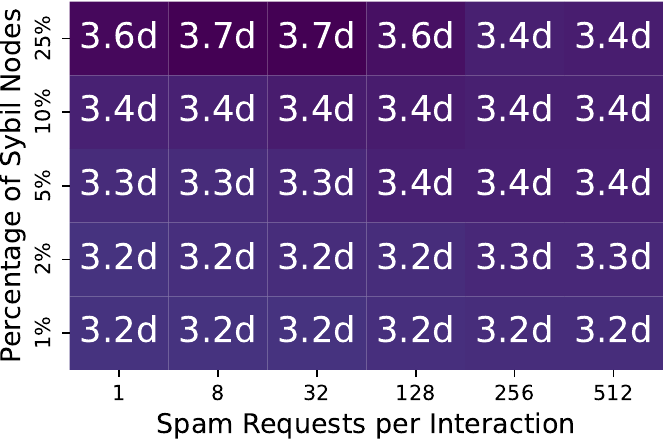}
    
  }    
  \caption{Metrics under epidemic routing, limited caching.}
    \label{fig:epidemic-metrics}
\end{figure}

\mypara{Epidemic routing suffers under adversary spam}
\Cref{fig:epidemic-metrics} presents simulation results where all pages are fetched via epidemic routing.
The plots vary the proportion of Sybil nodes and the number of spam requests each Sybil node makes per interaction.
In a benign scenario, epidemic routing achieves similar page request satisfaction but with far worse latency: $90^{th}$ percentile latency is 1.1 days for \CTTF compared to 3.2 days for epidemic routing.
Epidemic routing is also more vulnerable to adversarial influence: adversaries request useless pages which fill the forwarding buffers of non-adversaries with spam.
With sufficient spam requests, just 1\% Sybil nodes can reduce request satisfaction to 49\%, lower than the 54\% request satisfaction seen in the harsher 25\% Sybil + 2,500 $\text{km}^2$ jamming scenario.

\mypara{\CTTF is also effective with random movement}
\wpes{When using the random movement model instead of the YJMob100K dataset, there is a clear relationship between adversarial behaviour and latency: 90$^{th}$ percentile latency increases from 5.5 hours in a benign scenario to 1 day with the most aggressive adversarial behaviour.
Page request satisfaction is less obviously correlated, with all request satisfaction rates between 62 and 66\%.
The random movement of users tends to concentrate them uniformly throughout the grid, meaning leechers are more likely to receive a balanced set of ratings from both adversaries and seeders, and the jamming of cells is less effective at preventing user interaction.}

\begin{figure}
    \centering
    \includegraphics[width=0.7\linewidth]{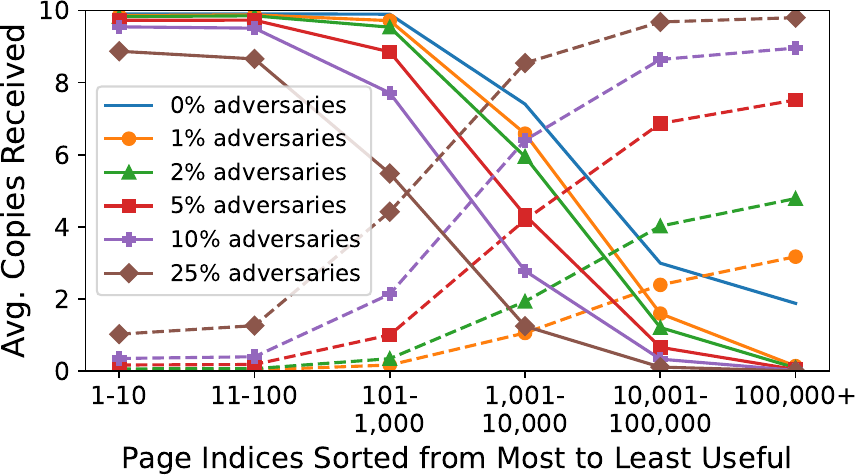}
    \caption{Solid lines: avg. page copies received from benign nodes; dotted lines: received from adversaries.}
    \label{fig:majority-vote}
\end{figure}

\mypara{Majority-vote suffices for popular pages}
\Cref{fig:majority-vote} presents \CTTF's behaviour when proxy downtime and missing page signatures causes users to rely on a majority-vote.
Each legitimate user attempts to fetch 10 copies of each requested page, while adversaries provide misinformation responses for all page requests.
For small percentages of Sybil nodes ($\leq 5\%$), on average more copies are fetched from non-adversaries than adversaries for the top 10\,000 pages.
However, if Sybil nodes are too numerous or if users request more obscure pages, they are more likely to be fed misinformation.
Still, \CTTF delivers legitimate content for popular pages under this extreme scenario.

\subsection{Transmission Performance and Efficiency}
\label{subsec:evaluation-microbenchmarks}
 To conduct microbenchmarks, we implemented a \CTTF prototype in 1\,300 lines of Kotlin and tested on a Samsung Galaxy S24 FE (Exynos 2400e CPU, 8GB RAM) and a ZTE Blade V40 Smart (Unisoc ums9230 CPU, 3GB RAM).

\mypara{Transmission speeds suffice for rich pages}
To measure Bluetooth classic transmission speeds, we exchanged 1MB of random data 100 times.
The transfers took $7.78s$ on average ($\sigma = 0.34s$), yielding a transfer rate of 129KB/s, which is in-line with previous estimates \cite{kamali2024anix,perry2022strong}.
Since the median mobile page weight is 2.5MB \cite{HTTPArchive}, this validates our previous assumption that 3-4 pages may be transferred within one minute of connection time.
\wpes{Similarly, we measured the throughput for BLE CoC, for users who are willing to sacrifice speed for additional privacy via randomized MAC addresses.
These transfers took $28.9s$ on average ($\sigma = 2.95s$), yielding a transfer rate of 35KB/s.
These results suggest that users who opt for using BLE CoC will need longer contact times to receive pages in full.}

\mypara{\CTTF does not consume excessive battery}
We used \CTTF to continuously alternate between exchanging 1MB messages and computing 16-bit proofs-of-work between two devices for one hour.
72 exchanges were able to be completed in this time period, which is roughly three times the average number of daily contacts (22.5) in the YJMob100K dataset.
The Samsung Galaxy discharged 304 mAh of its 4700 mAh battery ($6\%$) while the ZTE Blade discharged 403 mAh of its 6000 mAh battery ($7\%$).
While this estimate does not account for continuous Bluetooth scanning, we reason that proof-of-work computation would be the primary power consumption bottleneck and the microbenchmark indicates this is not prohibitive to the app's deployment.

\mypara{Proof-of-work computation times vary acceptably across devices}
While the Samsung Galaxy is a much more computationally powerful device compared to the ZTE Blade, a 17-18 bit proof-of-work difficulty produces reasonable delays on both devices (17-bit - ZTE: 30.3s avg., SG: 16.6s; 18-bit - ZTE: 74.2s, SG: 29.7s).
This experiment indicates that a well-chosen difficulty does not excessively burden low-end devices nor trivializes the proof-of-work for high-end ones.

\section{Related Work}

\mypara{Accessing digital content with limited connectivity}
Existing work has explored how to maintain access to digital content during limited or disrupted connectivity. %
We now describe some of these approaches, which, despite varying in scope, share the goal of improving access to information when Internet infrastructure is unavailable or unreliable.

\myparait{Decentralized browsers}
Ceno \cite{ceno} allows users to fetch webpages through proxies and cache them for others.
However, since cached webpages are fetched from other users via uTorrent, Ceno cannot supply webpages during blackouts.
\wpes{Ceno also does not replicate popular webpages across devices; local sharing of a webpage through Ceno would require finding a user who requested the webpage before the blackout and connecting over LAN, compared to \CTTF's automated caching, discovery, and retrieval.}

\myparait{Ad-hoc data delivery in disasters and underserved regions}
Previous systems have explored serving digital contents to regions that are disconnected by natural disasters or communication infrastructure failures.
Examples include the use of drones~\cite{cbc-facebook-solar-drone}, stratosphere balloons~\cite{loon}, satellite broadcasts~\cite{knapsackforhope,othernet,outernet}, and community networks~\cite{guifi}.
While these systems sustain basic connectivity or offer bulk content dissemination, they rely on centralized broadcast models, specialized hardware, or lack content-aware coordination.
Our work complements these efforts by focusing on local, verifiable caching and cooperation among users without external infrastructure dependencies during a blackout.

\myparait{Archiving tools}
Offline archiving tools allow users to store and browse collections of static content.
Such tools exist for Wikipedia~\cite{kiwix,wikitaxi} and for generic websites~\cite{archivebox,webrecorder}, enabling users to capture and replay web pages offline.
These tools aim to the preserve content in the long-term, operate in a per-user fashion, and lack support for adaptive caching strategies or coordination among multiple users. 
In contrast, \CTTF supports dynamic caching and sharing driven by users' perceptions on web content's utility.

\myparait{Read-it-later applications}
Instapaper~\cite{instapaper} and Raindrop~\cite{raindrop} allow users to save articles for later consumption, and are designed for convenience in intermittent connectivity scenarios. However, they require explicit user action to save content and offer no guarantees of authenticity when sharing pages across users. \CTTF builds on similar offline-first principles but extends them to support automated, collaborative caching while allowing content verification.

\mypara{Verifying page provenance}
\CTTF uses a trusted suite of proxies to verify that pages were authentically fetched and not tampered with.
We briefly summarize alternative approaches to proving and verifying the contents and origins of pages.

\myparait{TLS extensions} Prior works~\cite{ritzdorf2017tls,tlssign,tlsevidence} have modified TLS to allow for verification; TLS-N~\cite{ritzdorf2017tls} generates non-interactive proofs about session contents for third-party verification.
Such solutions require server-side adoption, which limits their utility in \CTTF's scenario where users may cache pages from a variety of sources. 

\myparait{TLS oracles} In contrast, TLS oracles attempt to prove data provenance without server modifications.
DECO~\cite{zhang2020deco}, Janus~\cite{lauinger2025janus}, and ORIGO~\cite{ernstberger2025origo} employ a three-party handshake with a TLS server, later allowing a prover to prove statements to a verifier about the TLS session data.
TLS oracles are unsuitable for \CTTF, as users lack prior knowledge of which peers they will contact, and the computational/communication overhead exceeds smartphone capabilities.

\section{Conclusion}
\label{sec:conclusion}

We introduced Cache to the Future: a distributed webpage archive for internet blackouts.
\CTTF explores a new scenario for blackout-resistant technologies, providing information access instead of messaging services.
We evaluated \CTTF through extensive simulations with real-world data, contrasting earlier works that model city-scale mobility through random movement on a small grid.%

\mypara{Future work}
More can be done to resist user fingerprinting and information leakage within \CTTF: \wpes{while BLE CoC allow for MAC address randomization, peers can directly observe non-fuzzed page ratings and requests.}
Differential privacy techniques have the potential to deter rating fingerprinting, while private information retrieval (PIR) schemes could obscure which pages are requested.
Unfortunately, existing PIR schemes~\cite{menon2022spiral,angel2018pir} impose too much computational/communication overhead to be directly applicable.
Investigating weaker privacy guarantees that may allow such schemes to run on smartphones is a promising avenue of future work.

\begin{acks} 
This work was supported in part by NSERC under grant RGPIN-2023-03304, and benefited from the use of the Cybersecurity and Privacy Institute (CPI) Chippie facility at the University of Waterloo. Our experiments were also supported by GCP infrastructure through the Google Cloud Research Credits program.
\end{acks}

\bibliographystyle{splncs04}
\bibliography{citations}

\appendix
\section{Additional Evaluation Results}
\label{appendix:additional-eval}

\begin{figure}
  \subfloat[Req. satisfaction (after a week). \label{fig:grid-satisfaction}]{
    \includegraphics[width=0.45\linewidth]{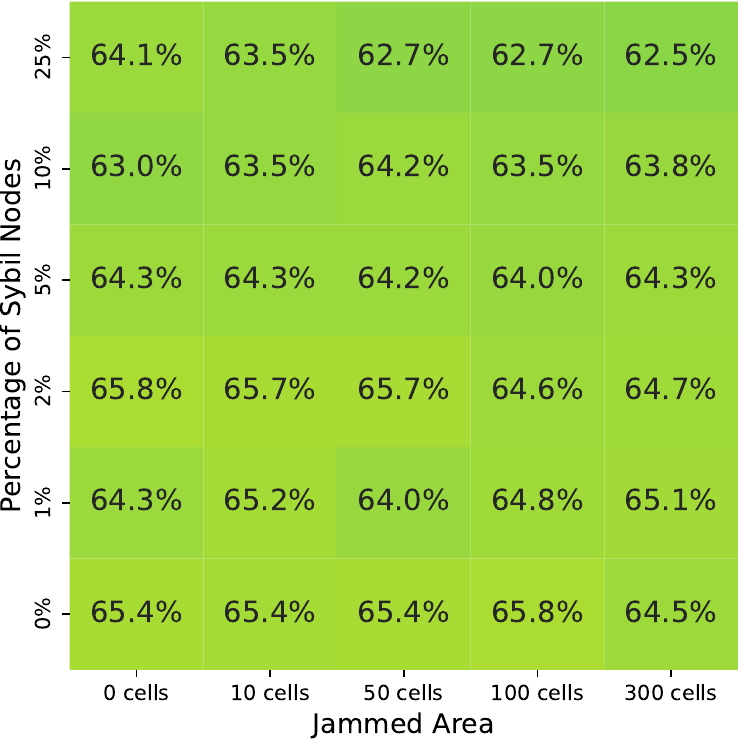}
  }
  \hfill
  \subfloat[90$^{th}$ percentile latency when requesting top-10k most useful pages. \label{fig:grid-latency}]{
    \includegraphics[width=0.45\linewidth]{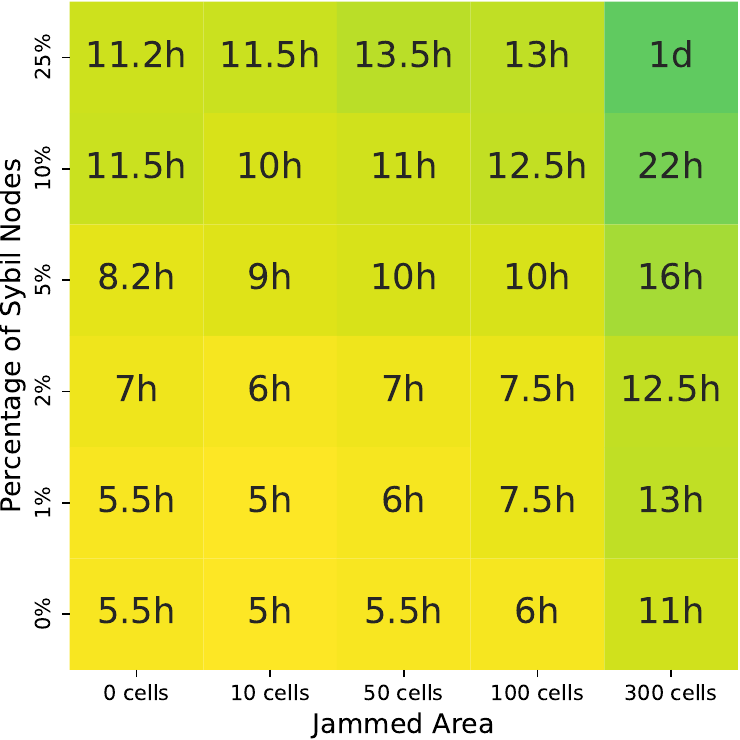}
    
  }    
  \vspace{-0.3cm}
  \caption{\CTTF metrics w/ jamming and Sybils; grid model.}
    \label{fig:grid-metrics}
    \vspace{-0.5cm}
\end{figure}

\mypara{\CTTF functions with randomized movement}
\Cref{fig:grid-metrics} presents \CTTF's performance in the random movement grid model used by Anix \cite{kamali2024anix} and ASMesh \cite{bienstock2023asmesh}.
These performance results are not directly comparable to these applications due to the different communication models, nor are they directly comparable to \CTTF's performance using the YJMob100K dataset; the random movement grid assumes only 600 users in the simulation, which is less than 2.5\% of the user count of the YJMob100K dataset.
While the random movement grid model has overall ``better'' latency than our simulations using the YJMob100K dataset, users move constantly under this model, whereas user movement is noticeably reduced during nighttime in the YJMob100K dataset.
Nevertheless, request satisfaction and latency indicate that \CTTF is usable in the random movement grid model even assuming adversarial influence via Sybil nodes and via jamming.

\begin{figure}
    \centering
    \includegraphics[width=0.75\linewidth]{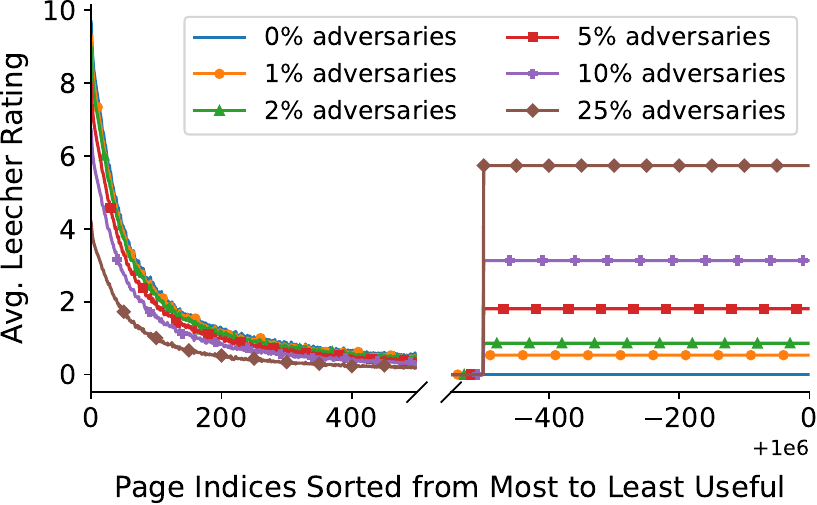}
    \vspace{-0.3cm}
    \caption{Avg. leecher ratings at different ratios of Sybil nodes.}
    \label{fig:average-leecher-ratings}
    \vspace{-0.2cm}
\end{figure}

\mypara{Rating manipulation and cache manipulation are correlated}
\Cref{fig:average-leecher-ratings} is the counterpart to \Cref{fig:pages-cached-by-leechers} in \S\ref{subsec:eval-caching}; this figure presents how leecher ratings are manipulated by Sybil nodes whereas the previous figure visualizes which pages are cached by leechers in the presence of Sybil nodes.
Observing these plots in tandem demonstrates that coordinated adversaries making up a significant portion of the population can manipulate leecher ratings.
This rating manipulation results in fewer leechers replicating copies of useful pages, and more leechers replicating copies of useless pages.

\begin{figure}
    \centering
    \includegraphics[width=0.75\linewidth]{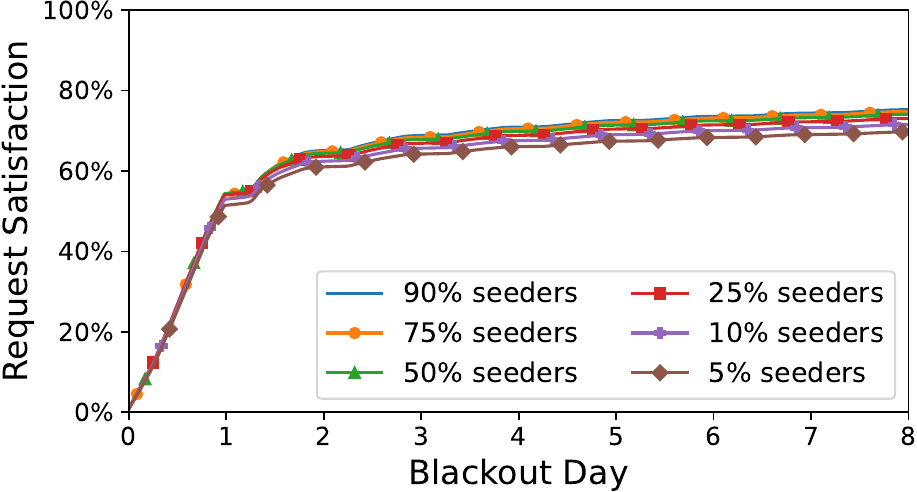}
    \vspace{-0.3cm}
    \caption{Req. satisfaction for different seeder/leecher ratios.}
    \label{fig:satisfaction-varySeeders}
    \vspace{-0.2cm}
\end{figure}

\mypara{In a benign scenario, few seeders are necessary to populate the cache}
\Cref{fig:satisfaction-varySeeders} presents the usual simulation with 1-week of pre-blackout rating exchanges and 8 days of blackout behaviour as we vary the ratio between leechers and seeders.
In this scenario with no adversaries, seeders are able to successfully populate leecher caches even when making up just 5\% of the total population: request satisfaction only drops by 6\% when comparing the 90\% seeder scenario to the 5\% seeder scenario.
Since the number of seeders primarily effects the caching behaviour, latencies did not vary significantly across different leecher/seeder ratios.
While \CTTF still functions well with few seeders in a benign scenario, we reason that a healthy population of seeders is still necessary to offset adversarial cache manipulation in a hostile scenario.

\begin{figure}
    \centering
    \includegraphics[width=0.75\linewidth]{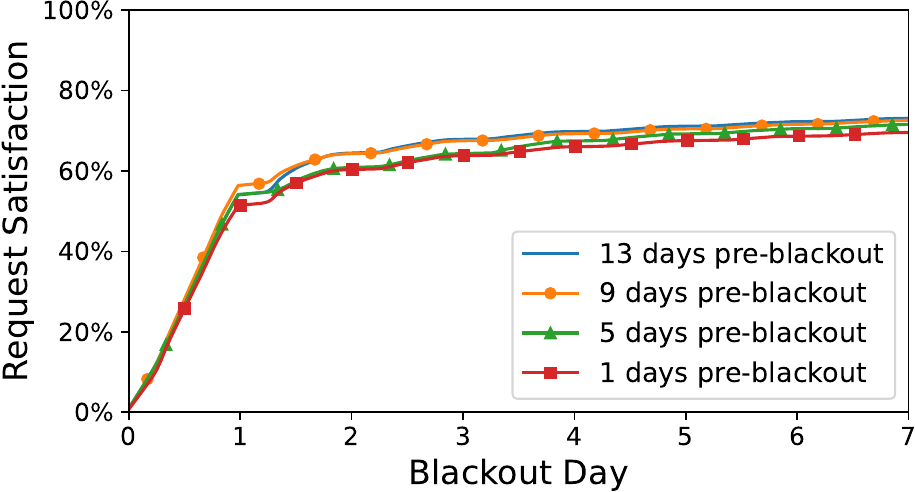}
    \vspace{-0.3cm}
    \caption{Req. satisfaction for different numbers of pre-blackout days.}
    \label{fig:satisfaction-cacheStartUp}
    \vspace{-0.2cm}
\end{figure}

\mypara{In a benign scenario, leecher caches can be populated quickly}
In our baseline we always simulate 7 days pre-blackout for leechers to compute local rating averages; \Cref{fig:satisfaction-cacheStartUp} presents request satisfaction curves for a variety of pre-blackout durations.
As shown in the plot, request satisfaction does not vary significantly given more pre-blackout days.
There is only a 4\% increase in request satisfaction with 13 days of pre-blackout rating exchanges as compared to a single day of pre-blackout rating exchanges.
This result indicates that leecher ratings become effective after a brief start-up period.
In an adversarial scenario, however, longer pre-blackout periods may be necessary to reduce variance; an unlucky leecher could encounter primarily adversaries early in the simulation which could skew ratings that would otherwise average out over a longer timeframe.

\begin{figure}
    \centering
    \includegraphics[width=0.75\linewidth]{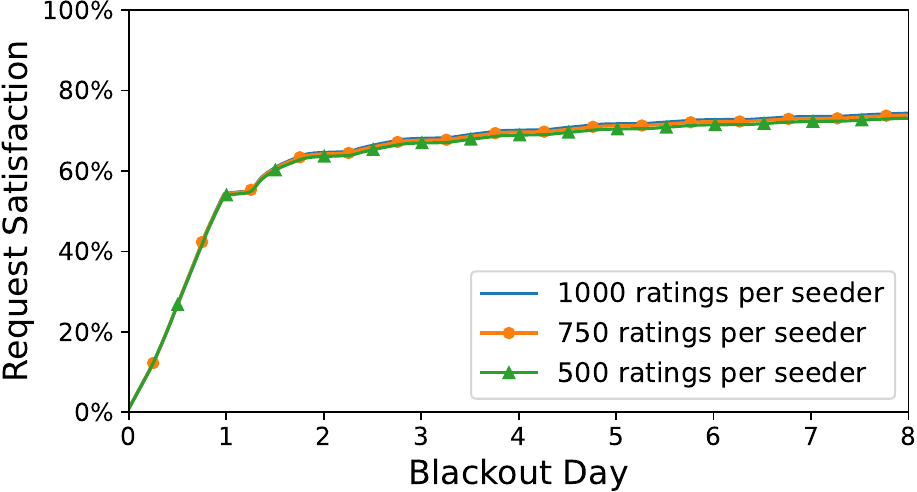}
    \vspace{-0.3cm}
    \caption{Req. satisfaction for different numbers of seeder ratings.}
    \label{fig:satisfaction-varyNumRatings}
    \vspace{-0.2cm}
\end{figure}

\mypara{In a benign scenario, seeders rating more pages does not impact results}
In the baseline simulation each seeder is assumed to rate 500 pages; these ratings are distributed to other users throughout the pre-blackout phase.
\Cref{fig:satisfaction-varyNumRatings} presents our baseline scenario modified such that seeders rate 500, 750, or 1\,000 pages each.
Neither request satisfaction nor latency are impacted by increasing the number of ratings each seeder distributes; this result implies that in a benign scenario, leecher ratings mimic ground truth ratings even with fewer page ratings distributed per-seeder.
In an adversarial scenario, increasing the number of ratings each seeder distributes may be advantageous in helping combat adversaries who always distribute the maximum number of spam ratings possible.

\section{Simulation Parameters}
\label{appendix:simulation-parameters}
\Cref{table:parameters} contains a list of parameters for the simulation testbed described in \S\ref{subsec:simulator-param-configs}. 
\clearpage

\begin{table*}[t]
\centering
\small
\caption{Simulation parameters used in \CTTF's evaluation. Bold values correspond to the configurations used in our baseline.}
\label{table:parameters}
\resizebox{0.99\textwidth}{!}{%
\begin{tabularx}{\textwidth}{l | X | X | X}
\textbf{Parameter} & \textbf{Description} & \textbf{Value} & \textbf{Rationale} \\
\hline
\addlinespace

\multicolumn{4}{c}{\textbf{Users' Distribution Parameters}} \\
\addlinespace

\texttt{SEEDER\_PERCENT} & Percent of seeder users in the simulation & 5\%, 10\%, \textbf{25\%}, 50\%, 75\%, 90\% & Conservative estimate \cite{meulpolder2010public,cox2010seeders} \\
\texttt{LEECHER\_PERCENT} & Percent of leecher users in the simulation & 5\%, 10\%, 25\%, 50\%, \textbf{75\%}, 90\% & Conservative estimate \cite{meulpolder2010public,cox2010seeders} \\
\texttt{ADVERSARY\_PERCENT} & Percent of adversaries in the simulation & \textbf{0\%}, 1\%, 2\%, 5\%, 10\%, 25\% & Increasingly stringent -- Stasi used 2\% of citizens as informants \cite{lerner2016rangzen} \\

\addlinespace
\multicolumn{4}{c}{\textbf{Page Rating Assumptions Parameters}} \\
\addlinespace

\texttt{PAGE\_COUNT} & Number of pages in the simulation universe & 1 million & Top million sites capture over 95\% of all page loads~\cite{ruth2022world} \\
\texttt{PAGES\_STORED} & Number of pages that can be cached on one device & 1\,200 & 3GB storage / 2.5MB median mobile page weight~\cite{HTTPArchive} \\
\texttt{PAGES\_RATED} & Number of unique pages each proactive user has rated pre-blackout & \textbf{500}, 750, 1\,000 & Near average page visits per day~\cite{crichton2021home} \\
\texttt{RATING\_CAP} & Number of ratings any user can transmit to another & 1\,000 & Mitigates adversarial manipulation \\
\texttt{INDIVIDUAL\_NOISE} & Additive noise added to each seeders ratings (exponentially decays) & 0.5 & Accounts for differences in ratings \\

\addlinespace
\multicolumn{4}{c}{\textbf{Page Request Frequency Parameters}} \\
\addlinespace

\texttt{PAGE\_REQUEST\_PROBABILITY} & Probability an individual requests a page at any given timestep & 0.25 & Average one request per 2 hours \\
\texttt{CONTACT\_PROBABILITY} & Probability two users will connect when in the same cell for a timestep & \textbf{100\%}, 75\%, 50\%, 25\%, 10\%, 5\% & Prior works \cite{pradeep2022moby,kamali2024anix,bienstock2023asmesh} fix at 100\% during evaluation \\
\texttt{FORWARDING\_LIMIT} & How many pages can be exchanged during communication & 4 & One minute of connection time allows approximately 10MB of data to be exchanged (\S\ref{subsec:evaluation-microbenchmarks}) \\

\addlinespace
\multicolumn{4}{c}{\textbf{Grid-Based Parameters}} \\
\addlinespace

\texttt{GRID\_SIZE} & Dimensions of the simulated world in grid cells & $200 \times 200$ & YJMOB100K~\cite{yabe2024yjmob100k} \\
\texttt{TOTAL\_USERS} & Total number of users in the simulation & 25\,000 & YJMOB100K~\cite{yabe2024yjmob100k} \\

\addlinespace
\multicolumn{4}{c}{\textbf{Adversary Behaviour Parameters}} \\
\addlinespace

\texttt{ADVERSARY\_STRATEGY} & Strategy of adversary for its ratings & Decrease top 500, Increase bottom 500 & Impede useful page caching and promote useless page caching \\
\texttt{TOP\_K\_JAMMING\_LOCATIONS} & How many of the most popular cells will adversaries jam & \textbf{0}, 10, 100, 1\,000, 10\,000 & Increasingly stringent \\

\addlinespace
\multicolumn{4}{c}{\textbf{Epidemic Routing Parameters}} \\
\addlinespace

\texttt{SPACE\_FOR\_FORWARDING} & Space dedicated to storing and forwarding non-personally-requested pages & 300 & 750MB with 2.5MB median mobile page weight~\cite{HTTPArchive} \\
\texttt{ADVERSARY\_FORCE\_MULTIPLIER} & Number of useless page requests adversaries make and forward & \textbf{N/A}, 1, 8, 32, 128, 256, 512 & Increasing spam \\

\addlinespace
\multicolumn{4}{c}{\textbf{Proof-of-Work Parameters}} \\
\addlinespace

\texttt{STALKING} & Whether adversaries engage in stalking for rating manipulation, necessitating PoW mitigation & \textbf{False}, True & Stalking and rapidly spoofing Bluetooth MAC address is an extreme threat model \\
\texttt{POW\_LIMIT} & Avg. time to compute the PoW, which limits how many rating exchanges can happen in one timestep & \textbf{N/A}, 1 second, 30 seconds, 1 minute & Investigating appropriate parameters \\

\end{tabularx}}
\end{table*}

\end{document}